\documentclass[%
 reprint,
 amsmath,amssymb,
 aps,soul,
pre,
]{revtex4-2}

\usepackage{graphicx}
\usepackage{dcolumn}
\usepackage{bm}
\usepackage{xcolor}
\usepackage{booktabs}

\renewcommand{\selectlanguage}[1]{}
\usepackage{lipsum}
\makeatletter
\newcommand*{\balancecolsandclearpage}{%
  \close@column@grid
  \cleardoublepage
  \twocolumngrid
}
\begin{document}

\preprint{APS/123-QED}

\title{Renormalised hydrodynamics in polar chiral active matter: \\ Spectral scaling and disorder-driven vortex clustering \\ in phase-coupled, motile oscillators}


\author{Magnus F Ivarsen}
\altaffiliation[Also at ]{
The European Space Agency Centre for Earth Observation, Frascati, Italy}
\email{Contact: magnus.fagernes@gmail.com}
\affiliation{Department of Physics and Engineering Physics, University of Saskatchewan, Saskatoon, Canada}%

\begin{abstract}
Active turbulence in overdamped chiral systems presents a complex challenge, namely the frequent exhibition of non-universal spectral scaling, creating large-scale coherent structuring that seemingly defies standard inertial fluid descriptions. In this study, we investigate the hydrodynamic limit of a two-dimensional polar chiral active fluid modeled as an ensemble of locally coupled, motile Kuramoto-Sakaguchi oscillators. By introducing a Renormalised Fluid Element (RFE) operator, we coarse-grain microscopic phase singularities, and in so doing, we isolate the macroscopic transport dynamics. We demonstrate that while the raw particle distributions consistently exhibit steep, dissipative energy spectra, associated with enstrophy injection at the microscale, the RFE-filtered field reveals a dual behaviour characterized by an inverse energy cascade. When the intrinsic frequency dispersion, drawn from a scale-free power-law distribution, is broad enough to seed fluctuations at all resolved scales, this hidden cascade acts akin to a topological heat pump, driving the system toward a state of macroscopic vortex clustering, structurally analogous to supersonic shallow water dynamics. Conversely, a narrow frequency dispersion results in kinetic arrest, forming an active vortex glass. These results suggest that overdamped phase-slaved active matter can sustain effective inertial cascades, providing a mathematical framework for understanding scale-dependent energy transport in driven chiral systems.
\end{abstract}




\maketitle


\section{Introduction}

Active turbulence, the chaotic spatiotemporal dynamics observed in bacterial
swarms \cite{wensink_meso-scale_2012,aranson_bacterial_2022}, cytoskeletal
extracts \cite{alert_active_2022}, and synthetic colloidal flocks
\cite{bourgoin_kolmogorovian_2020,qi_emergence_2022}, arises in the overdamped
limit ($\mathcal{R}e \approx 0$): the energy sustaining the chaotic transport is
injected continuously at the microscale, not supplied by an inertial cascade
\cite{saintillan_active_2013,bhattacharjee_activity_2022}.

Observations nevertheless reveal large-scale coherent structures reminiscent of
inertial flows \cite{urzay_multi-scale_2017}, and the measured spectral
exponents are conspicuously non-universal \cite{alert_active_2022}. Theories
built on the hydrodynamics of active nematics \cite{doostmohammadi_active_2018}
predict steep spectra ($E(k) \propto k^{-4}$ or $k^{-3}$) dominated by defect
proliferation \cite{bratanov_new_2015,giomi_geometry_2015}, and such spectra are
indeed widely observed
\cite{toner_flocks_1998,wensink_emergent_2012,giomi_defect_2013}. They do not,
however, account for the inverse energy cascades ($k^{-5/3}$) reported in chiral
active fluids
\cite{meckeEmergentPhenomenaChiral2024,reevesEmergenceLanesTurbulentlike2021}
and in two-dimensional quantum turbulence \cite{reeves_inverse_2013}.

Chiral active matter breaks detailed balance through intrinsic rotation
\cite{markovich_chiral_2025}, with known consequences for transport, such as odd
viscosity \cite{marmol_colloquium_2024,maitra_activity_2025} and entropy-driven
ordering \cite{vijayan_disorder_2025}. Its thermodynamic end-state is less
settled: does sustained microscopic rotation feed a chaotic steady state, or
does it select a specific macroscopic attractor?

We address this question in a minimal model: $N$ motile oscillators whose
velocities are slaved to an internal phase, coupled through a local
Kuramoto-Sakaguchi interaction \cite{ivarsenPolarChiralActive2026}. The sole
energy source is quenched frustration, intrinsic frequencies $\omega_i$ drawn
from a scale-free power-law distribution, which injects enstrophy at the
microscale across all resolved scales. In a companion paper
\cite{ivarsenPolarChiralActive2026} we established that the agent dynamics are
isomorphic to a disordered, resistively shunted Josephson array; here we ask
what hydrodynamics this array supports. To that end we introduce a Renormalised
Fluid Element (RFE) operator, a coarse-graining over the interaction timescale
that removes the slaved microscopic fluctuations and their phase singularities.

Three results follow. First, the system exhibits a spectral duality: the raw
particle field shows a steep, enstrophy-dominated spectrum, while the RFE field
carries a negative spectral flux with a spectrum consistent with $k^{-5/3}$.
Second, the frustration distribution $\Delta\omega$ controls a three-phase
diagram, spanning global synchronisation, an arrested vortex glass, and a
vortex condensate. Third, in the condensate phase the inverse transfer drives
the system toward a large-scale dipole that structurally mimics Onsager's
point-vortex condensate \cite{onsager_statistical_1949}, maintained as a
dynamic attractor rather than a thermodynamic equilibrium. Filters analogous
to the RFE may recover comparable hidden transport in other overdamped active
systems, a possibility we return to in the Discussion.

\section{Methodology}

\subsection{The Model}

We define the system as a polar chiral active fluid composed of $N=50,000$ particles, or discrete, phase-slaved agents, in a bounded periodic domain $\mathcal{D} \in \mathbb{R}^2$. The state of the $i^\text{th}$ agent is defined by its position $\mathbf{r}_i(t)$ and its internal phase $\phi_i(t) \in [0, 2\pi)$. The agents do not interact via pairwise collisions (as in granular matter) but through a local mean field. Following Ref.~\cite{ivarsenPolarChiralActive2026}, we define the complex order parameter field $Z(\mathbf{r}, t)$ as the convolution of the microscopic agent distribution with a finite-range interaction kernel $G$:
\begin{multline} \label{eq:kernel}
    Z(\mathbf{r}, t) = R(\mathbf{r}, t) e^{i\Psi(\mathbf{r}, t)} = \\ = \int_{\mathcal{D}} G(|\mathbf{r} - \mathbf{r}'|) \left[ \sum_{j=1}^N \delta(\mathbf{r}' - \mathbf{r}_j(t)) e^{i\phi_j(t)} \right] d\mathbf{r}',
\end{multline}
where $G(|\mathbf{r}-\mathbf{r}'|)$ is the Green's function of the interaction (modeled as a Gaussian with kernel size $\sigma$), $R(\mathbf{r}, t)$ is the local phase coherence (or order parameter), where $R \approx 1$ implies high local synchronisation and $R \approx 0$ corresponds to a phase singularity, or defect core. $\Psi(\mathbf{r}, t)$ is the local mean phase. The convolution kernel $G$ is interpreted in the context of chemical gradients or electric fields, which govern the dense active matter interactions (e.g., bacterial swarms). By modifying $G$, we can potentially simulate anisotropic media (like magnetized plasmas) without changing the agent logic. We note that chirality here is a property of the individual agents; the sense in which the macroscopic ensemble breaks mirror symmetry is addressed in Sec.~\ref{sec:constraints}.

\begin{figure}
    \centering
    \includegraphics[width=0.5\textwidth]{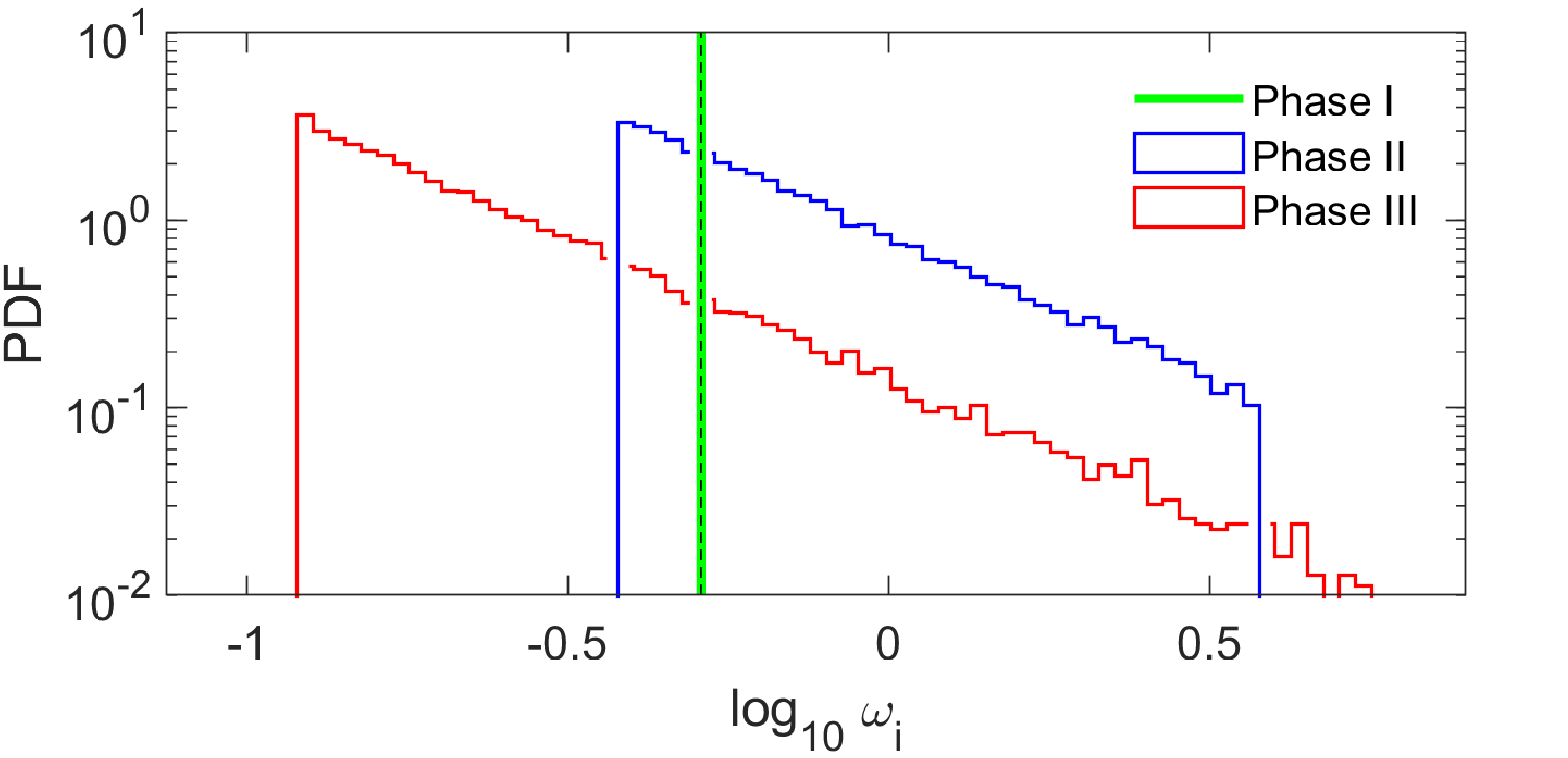}
    \caption{The three distributions $\Delta \omega$, corresponding to intrinsic frustration, or noise, i.e.,  the values of $\omega_i$. The three distributions, colored red, blue, and green, correspond to the three thermodynamic phases in Table~\ref{tab:phase_diagram} and Figure~\ref{fig:ensemble}. The lower limit on the $x$-axis is given by $(2\Lambda)^{-1}$, the hypothetical ``box-size fluctuation.'' We rely on the ``Phase III'' distribution for most of our numerical simulations, with the two remaining distributions used to delineate the validity of our results. Note that the plot shows only absolute values ($\omega_i$ is symmetric around 0). See Ref.~\cite{ivarsenPolarChiralActive2026} for details on this distribution.}
    \label{fig:distro}
\end{figure}

Following Ref.~\cite{ivarsenPolarChiralActive2026}, the internal phase $\phi_i$ evolves according to an overdamped, localized, driven Kuramoto-Sakaguchi-like interaction \cite{acebron_kuramoto_2005}. In a companion paper, Ref.~\cite{ivarsenPolarChiralActive2026}, we identify this interaction as isomorphic to the \textit{Adler equation} \cite{wiesenfeld_new_1996,danner_injection_2021} for disordered Josephson junction arrays,
\begin{equation} \label{eq:kuramoto}
    \dot{\phi}_i = \underbrace{\omega_i}_{\text{Driver}} + \underbrace{a_0 R(\mathbf{r}_i) \sin(\Psi(\mathbf{r}_i) - \phi_i)}_{\text{Synchronisation force}} + \underbrace{\eta_i(t)}_{\text{Noise}},
\end{equation}
where $\omega_i$ are natural frequencies drawn from a power-law distribution $P(\omega) \sim \omega^{-n}$, $a_0$ is the coupling strength, $\eta_i$ is a Gaussian white noise term (amplitude 0.15), and $n=1.5$ is a positive constant of order unity (see Ref.~\cite{ivarsenPolarChiralActive2026} for a study of the role of this parameter, and Figure~\ref{fig:distro} for the distributions in $\omega_i$ that we apply in the present papers). $R_i$, the local order parameter, ensures that synchronisation is triggered by initial random local alignments in phase. $\omega_i$ (and the stochastic term $\eta_i$) prevent a ``ferromagnetic collapse'', in which the system simply synchronises globally ($\phi_i = \text{const}$) causing a uniform ballistic drift in the agent ensemble. By forcing the particles to oscillate with an inherent frequency (or attempt to do so), the $\omega_i$ terms act as \textit{frustration}, constantly injecting enstrophy, or vorticity, at the micro-scale. The power-law form of $P(\omega)$ makes this frustration scale-free, supplying fluctuation power at all scales up to the box size; the near-box-size criterion of Section~\ref{sec:tridiscussion} is the operational consequence. As we shall demonstrate, this feeds an inverse energy cascade in the two-dimensional simulations.

The definitive equation of our model is the agent-based, non-inertial slave principle, where the velocity vector is explicitly determined by the instantaneous phase:
\begin{equation} \label{eq:velocity}
    \dot{\mathbf{r}}_i = \mathbf{v} =  v_0 (\cos \phi_i, \sin \phi_i),
\end{equation}
where $v_0$ is the active swim speed and  $\mathbf{r}_i=(x_i,y_i)$. This means that an agent's `oscillation' is in fact its motion, rendering the agent ensemble a collection of motile oscillators. The agents are therefore without conventional inertia: $\sum \mathbf{F} \neq m\mathbf{a}$; the system is fully overdamped, in a low Reynolds number limit \cite{dey_dynamic_2019}. 
The velocity field naturally forms vortices, introducing singularities,  
and so to define a valid continuum fluid, we introduce the renormalised fluid element (RFE), which we motivate theoretically in Appendix~\ref{app:hydro}.


At this point, we identify the characteristic scales in our model; the kernel size $\sigma$ and the active swim speed $v_0$, which yield the interaction time $\tau \equiv \sigma / v_0 = (3\text{ grid units})/(0.5\text{ grid units/s})=6$~s, yielding the dimensionless units $\tilde{x} = x/\sigma$, $\tilde{t} = t/\tau$, and $\tilde{\omega} = \omega (\sigma / v_0)$ (we subsequently drop the tilde). The system size becomes $\Lambda=2L/\sigma\approx13$. A natural definition of the RFE is then afforded by $\tau$; we define the RFE as the center of mass of a particle's trajectory during $\tau$, effectively coarse-graining the singularities:
\begin{equation} \label{eq:gc}
    r_i^\text{RFE} = \text{angle}\left( \frac{1}{\tau} \sum_{t'=t-\tau}^{t} \hat{r}_i(t') \right) \cdot \frac{\Lambda}{\pi},
\end{equation}
with $\hat{r}_i(t)$ being,
\begin{equation} \label{eq:gc2}
    \hat{r}_i(t) = \exp\left(i \cdot \frac{r_i(t)}{\Lambda} \cdot \pi \right),
\end{equation}
the complex phasor locations mapped to a unit circle (necessitated by the periodic boundary conditions). The transformation Eqs.~(\ref{eq:gc}, \ref{eq:gc2}) constitute the regularization of the topological defects in $\phi_i$, which become physical flow vortices in $\mathbf{r}_i$. 


The correlation wavenumber of the local mean field is $k_c=2\pi/\sigma$, or simply $2\pi$ with normalized units, establishing a natural scale separation that marks the transition between the microscopic active bath ($k > k_c$) and the macroscopic hydrodynamic limit ($k < k_c$), with $k$ being the normalized wavenumber. The RFE operator effectively acts as a low-pass filter with cut-off $k_c$. The kernel size $\sigma$ is thus a dynamical control parameter in its own right, setting the interaction range in Eq.~(\ref{eq:kernel}), the RFE averaging time $\tau$, and the cutoff $k_c$; Appendix~\ref{app:robustness} tests the robustness of the reported phase behaviour against variations in $\sigma$ at fixed $P(\omega)$.

\begin{figure*}
    \centering
    \includegraphics[width=\textwidth]{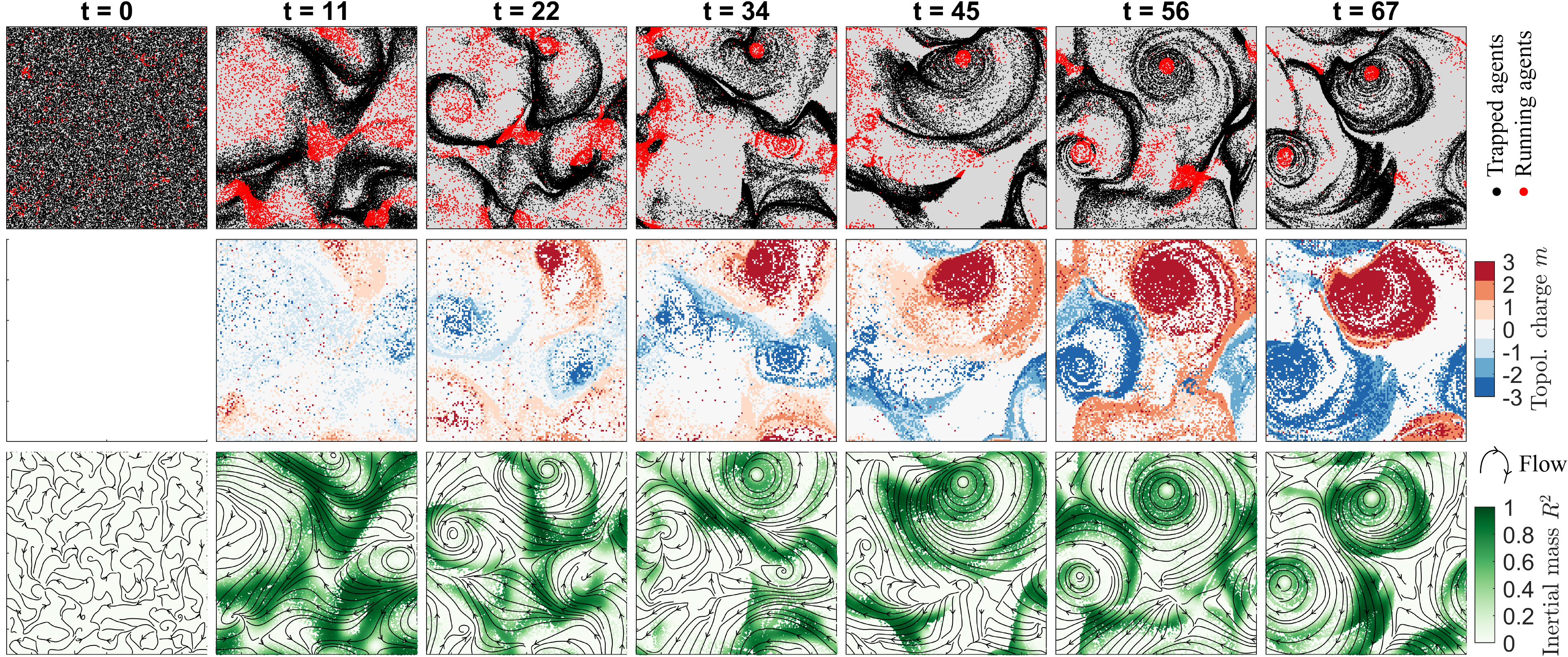}
    \caption{\textbf{The emergence of the inviscid dynamics in simulated polar chiral active matter,} shown in seven temporal snapshots of the simulation space. The top row shows the raw particle point-clouds, with black denoting `trapped' (superconducting) agents and red denoting `running' (phase slipping) agents \cite{ivarsenPolarChiralActive2026}, the middle row shows topological charge $m$, and the bottom row shows inertial mass $R^2$ (with flow lines overlaid), using the colorscales indicated.
    }
    \label{fig:0}
\end{figure*}

\subsection{Theoretical foundation}

To bridge the discrete dynamics and the continuum fluid, we present, in Appendix~\ref{app:hydro}, a lengthy derivation allowing us to assert with some confidence that the dynamics produced by the model equations (Eqs.~\ref{eq:kernel}--\ref{eq:gc2}) converge to an effective hydrodynamic limit described by shallow water hydrodynamics with topography. A brief summary follows.


By decomposing the particle velocity into a mean flow with fluctuations, we find that the macroscopic velocity $\mathbf{u}$ is governed by the mean phase gradient $\nabla \Psi$ scaled by the local order parameter $R$: $\mathbf{u} \sim R \nabla \Psi$. While we derive a set of constraining equations that are structurally equivalent to shallow water hydrodynamics, there are striking differences: the effective inertial mass in our model scales with local order, $\lambda = R^2$,
a relationship strictly predicted by the macroscopic \textit{phase stiffness} of the agent dynamics as a motile, disordered Josephson array \cite{ivarsenPolarChiralActive2026} (Eq.~\ref{eq:inertia3}). In other words, the renormalised fluid exhibits an effective mass constituted by the phase stiffness of the synchronised oscillators. 

How can inertial transport arise in a system whose constituents obey strictly first-order dynamics? The resolution is to embow the phase field as the momentum carrier. An isolated agent has no momentum memory, or inertia: its velocity is slaved to its instantaneous phase (Eq.~\ref{eq:velocity}). A synchronised patch, by contrast, does have memory. Redirecting the collective velocity of the patch requires rotating every constituent phase against the Kuramoto coupling torque, and the energetic cost of that collective rotation is the Ginzburg-Landau phase stiffness, which scales as $R^2$ (Eq.~\ref{eq:kappaid}). This stiffness endows the coarse-grained flow with resistance to changes in its velocity, which is an effective inertial mass in the operational sense: $\lambda = R^2$ (Eq.~\ref{eq:inertia3}). The RFE operator exposes this collective degree of freedom; averaging over the interaction time $\tau$ removes the fast, slaved fluctuations that dominate the raw field, leaving the slowly evolving momentum stored in the gradients of the mean phase $\Psi$. The situation is, by chance or design notwithstanding, analogous to a microelectrical device, namely the resistively shunted Josephson array, a voltage oscillator in which individually dissipative (running) agents support an information (phase rigidity) supercurrent carried by the collective phase itself \cite{ivarsenPolarChiralActive2026}.

A brief summary follows. From the foregoing identification of the phase pattern motion necesarily featuring an inertial component, we identify the active enthalpy $h_{\text{eff}}=v_0^2R^2/2$, which implies a sound speed $c_s \propto R$ (Eq.~\ref{eq:cs}). Since the flow velocity $\mathbf{u}$ scales identically (Eq.~\ref{eq:macro}), the agent ensemble becomes locked into a global supersonic state ($M\approx\sqrt{2}$, Eq.~\ref{eq:mach}). Consequently, phase defects become shielded by sonic horizons, as they are unable to escape. The ensemble thereby naturally phase-separates, effectively isolating agents whose intrinsic frustration precludes synchronisation. This prevents radiative decay and enforces strict conservation of topological charge.



%

As the mean field $\Psi(r)$ is the argument of a single-valued complex field (Eq.~\ref{eq:kernel}), its circulation along any closed loop $\partial\Sigma$ enclosing a defect core is necessarily quantized \cite{mermin_topological_1979,pismen_vortices_1999}, mirroring the winding number stability in discrete oscillator ensembles \cite{delabays_multistability_2016},
\begin{equation} \label{eq:winding}
    \oint_{\partial \Sigma} \nabla \Psi \cdot d\mathbf{l} = 2\pi m. \quad (m \in \mathbb{Z})
\end{equation}
This leads us to conjecture that the interaction energy of the ensemble is given by the Kirchhoff-Onsager Hamiltonian \cite{ovchinnikov_energy_2002,pelinovsky_variational_2011},
\begin{equation} \label{eq:hamiltonian}
    H = -\sum_{i \neq j} \Gamma_i \Gamma_j \ln |\mathbf{r}_i - \mathbf{r}_j|.
\end{equation}
While this assertion holds to close approximation (as we shall demonstrate, we observe what closely resembles inviscid turbulence), we caution that we are not able to conclude that the renormalised field  is strictly governed by a conservative Hamiltonian.

\subsection{A dynamic attractor}


In classical inviscid fluids on a bounded domain $\mathcal{D}$, the point-vortex gas model predicts that phase space volume $\Omega(E)$ eventually decreases at high energies, creating a negative-temperature regime ($T < 0$) that favors the merging of like-signed vortices into a large-scale dipole \cite{onsager_statistical_1949}. By applying the RFE operator, we observe that our renormalised fluid approaches a remarkably similar macroscopic end-state: robust vortex clustering. However, unlike a conservative Euler fluid, our active chiral model is driven and dissipative. The large-scale coherent structures are continuously eroded by supersonic shocks, which thermalise inertial energy back into the active bath. Consequently, the macroscopic dipole emerges not as a strict thermodynamic equilibrium, but as a \textit{dynamic attractor}. Here, the active bath constantly rebuilds topological order using newly thermalised energy, sustaining a steady-state that structurally mimics Onsager condensation without necessarily satisfying the rigorous statistical mechanical requirements of a true negative-temperature regime. Crucially, this dynamic clustering parallels observed inverse-energy cascades \cite{wensink_meso-scale_2012}, motility-induced phase separation \cite{cates_motility-induced_2015}, and entropy-driven ordering \cite{vijayan_disorder_2025} in active matter studies.






\subsection{Spectral scaling}

This framework predicts a spectral dichotomy based on the scale of observation. 
At the microscopic scale, the raw particle field is dominated by the singular geometry of the phase defects ($\nabla \phi \sim 1/r$), forcing the system into the enstrophy-dominated regime with a steep spectral index of approximately $E_{raw}(k) \propto k^{-8/3}$, reminiscent of  $E_{raw}(k) \propto k^{-3}$ \cite{kraichnan_inertial-range_1971,borue_spectral_1993,borue_inverse_1994,chertkov_dynamics_2007}. Conversely, at macroscopic scales, the transport dynamics closely approximate those described by the Kirchhoff-Onsager Hamiltonian (Eq. \ref{eq:hamiltonian}) for an effective inviscid fluid. Consequently, applying the RFE field to filter out microscopic singularities reveals an energy transfer signature consistent with an inverse cascade. Within the specific high-activity regime of our model, this spectrum approaches the classical Kolmogorov scaling $E_{RFE}(k) \sim k^{-5/3}$ \cite{kolmogorov_local_1968,reeves_inverse_2013}.


\begin{figure*}
    \centering
    \includegraphics[width=\textwidth]{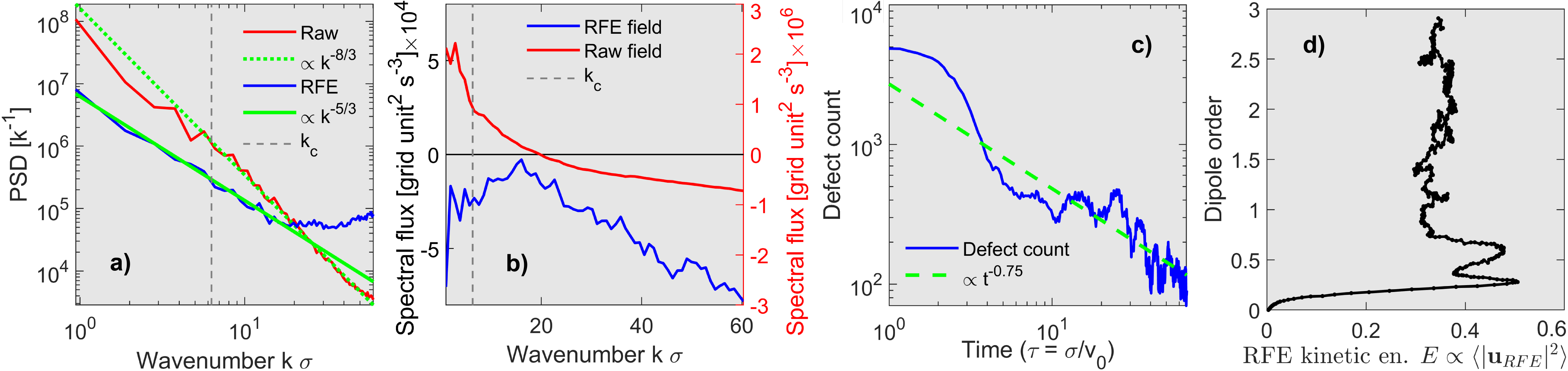}
    \caption{\textbf{Energy characteristics of the simulation run in Figure~\ref{fig:0}}. \textbf{Panel a)} shows the defect (phase singularity) count, with a $t^{-0.75}$ powerlaw scaling \cite{carnevale_evolution_1991} shown with a dashed, green line. \textbf{Panel b)} shows the spectral flux (Eq.~\ref{eq:spectralflux}) for the RFE (blue, left $y$-axis) and raw (red, right $y$-axis) fields, taking the median of the entire simulation. \textbf{Panel c)} shows the energy spectrum $E(k)$ for the RFE (blue) and raw (red) fields; a spectral scaling of $k^{-5/3}$ and $k^{-8/3}\approx k^{-2.67}$ is indicated with dotted and solid green lines respectively, and $k_c$ is indicated with a dashed, grey line, as in Panel b). \textbf{Panel d)} shows the thermodynamic trajectory, plotting order versus coarse-grained kinetic energy for the simulation evolution.}
    \label{fig:1}
\end{figure*}

We quantify the non-linear transfer of energy across scales using the standard spectral flux $\Pi(k)$, defined via the advective transport term $(\mathbf{v} \cdot \nabla)\mathbf{v}$ characteristic of fluid turbulence,
\begin{equation} \label{eq:spectralflux}
    \Pi(k) = -\int_0^k T(k') dk',
\end{equation}
where the energy transfer function $T(k)$ is defined for a velocity field $\mathbf{v}$ as \cite{pope_turbulent_2000},
\begin{equation}
    T(k) = \sum_{|\mathbf{q}|=k} \text{Re}\left[ \hat{\mathbf{v}}_\mathbf{q}^* \cdot \mathcal{F}\left\{ -(\mathbf{v} \cdot \nabla)\mathbf{v} \right\}_\mathbf{q} \right],
\end{equation}
which sums contributions of all discrete wave vectors $\mathbf{q} = (q_x, q_y)$ that fall within the annular shell where the magnitude $|\mathbf{q}| \approx k$, to the real part of the Fourier transform of the advective transport term, where $\hat{\mathbf{v}}_\mathbf{q}^*$ is the complex conjugate of the velocity field in Fourier space at wave vector $\mathbf{q}$. A broad negative plateau in $\Pi(k)$ would provide direct evidence of an inverse cascade.

\section{Results}

We have performed a series of numerical simulations of our model; we direct the reader to the Supplementary Materials online, where we provide three videos of the simulations, for detailed renderings.

The continuous simulation domain is defined as a bi-periodic square of dimensions $2L \times 2L$ (where $L$ is the physical half-width of the domain). While the $N=50,000$ agents evolve in continuous space via a standard Euler integration scheme, the macroscopic fields, including the RFE continuous field and the spectral flux calculations, are evaluated by projecting the discrete particle data onto a fixed Eulerian grid with a resolution of $128 \times 128$ grid points. The system is initialized with a random uniform distribution of phases and positions, an ``active soup'' state, and evolved for $T_{\max}=400$~s (20,000 time steps). The complex order parameter field $Z(\mathbf{r},t)$, and consequently the local coherence $R(\mathbf{r},t)$, were evaluated at each integration time step ($\Delta t = 0.02$), by binning the discrete agent distribution onto a $128\times128$ spatial grid, and the convolution with the Gaussian kernel $G$ is performed via two-dimensional Fast Fourier Transforms (FFT), allowing for continuous, real-time feedback between the mean field and the individual agents.

The temporal evolution of the system, shown in Figure~\ref{fig:0}, reveals a spontaneous transition from microscopic chaos ($t=0$) to macroscopic order ($t=11)$, as small-scale phase defects nucleate and annihilate rapidly. Surviving defects begin to cluster and merge, forming a sloshing, large-scale dipole ($t=22$ and beyond), where we direct the reader to the provided Video~S1 in the Supplementary Materials.

To discern the mechanism driving this dynamic condensation, we analysed the kinetic pathways of the topological defects (Figure~\ref{fig:1}a), the decay of the defect number density $N_d(t)$. The decay is consistent with the theoretical $t^{-0.75}$ scaling \cite{carnevale_evolution_1991,larichev_weakly_1991}: extended ensemble runs (16 members, evolved to $t = 2000$~s; Appendix~\ref{app:robustness}) yield a fitted exponent of $-0.776$ with a bootstrapped 95\% confidence interval of $[-0.804, -0.743]$, which contains $-3/4$ and excludes both the diffusive $t^{-1}$ rate and the compressible $t^{-2/3}$ rate.
The qualitative dominance of vortex merger, the kinetic pathway expected from dispersive shock dynamics \cite{hoefer_dispersive_2006, gurevich_nonstationary_1973,larichev_weakly_1991} and acoustic horizons \cite{unruh_experimental_1981} rather than from passive strain fields or random-walk annihilation \cite{shankar_topological_2022}, is nevertheless robust (Video~S1).

The directionality of the energy transfer is assessed with the spectral energy flux $\Pi(k)$ (Figure~\ref{fig:1}b), where the negative spectral flux ($\Pi(k) < 0$) at wavenumbers below $k_c$ indicates net upscale energy transfer; energy injected at the microscale by $\omega_i$ is transported toward the box size. We note that the negative flux region does not form a broad, constant plateau spanning the full interval over which the RFE spectrum approaches $k^{-5/3}$; the exponents reported in this study should accordingly be read as approximate, phenomenological descriptors rather than as established inertial-range scaling laws (Appendix~\ref{app:robustness} examines how the accessible scale separation varies with the kernel size $\sigma$). 


Figure~\ref{fig:1}c) shows that the raw particle velocity field exhibits a steep, dissipative effective spectrum, scaling approximately as $E_{raw}(k) \sim k^{-8/3}$ within the observed wavenumber range. This steep slope reflects the singular geometry of the topological defects: the raw agents are kinematically pinned to the sharp cusps of the phase field ($\nabla\phi \sim 1/r$). In the confined geometry of our two-dimensional simulation, this tight packing of vorticity filaments \cite{bradley_energy_2012}, here driven by the continuous injection of microscopic frustration ($\omega_i$), generates a highly dissipative active bath. The kernel-size sweep in Appendix~\ref{app:robustness} indicates that this spectrum steepens toward $k^{-3}$ as the scale separation is increased, identifying the $k^{-8/3}$ value as a finite-scale-separation result. However, the exact exponent of this microscopic bath is secondary to the primary spectral dichotomy observed when we apply the RFE operator to the ensemble. This filters out the microscopic singularities by averaging over the interaction scale $\tau$, revealing an approximately Kolmogorov-like range: the RFE spectrum $E_{RFE}(k)$  (the blue spectrum in Figure~\ref{fig:1}c) is consistent with the classical $k^{-5/3}$ slope characteristic of two-dimensional inviscid fluid turbulence over the observed wavenumber interval.

\begin{figure*}
    \centering
    \includegraphics[width=\textwidth]{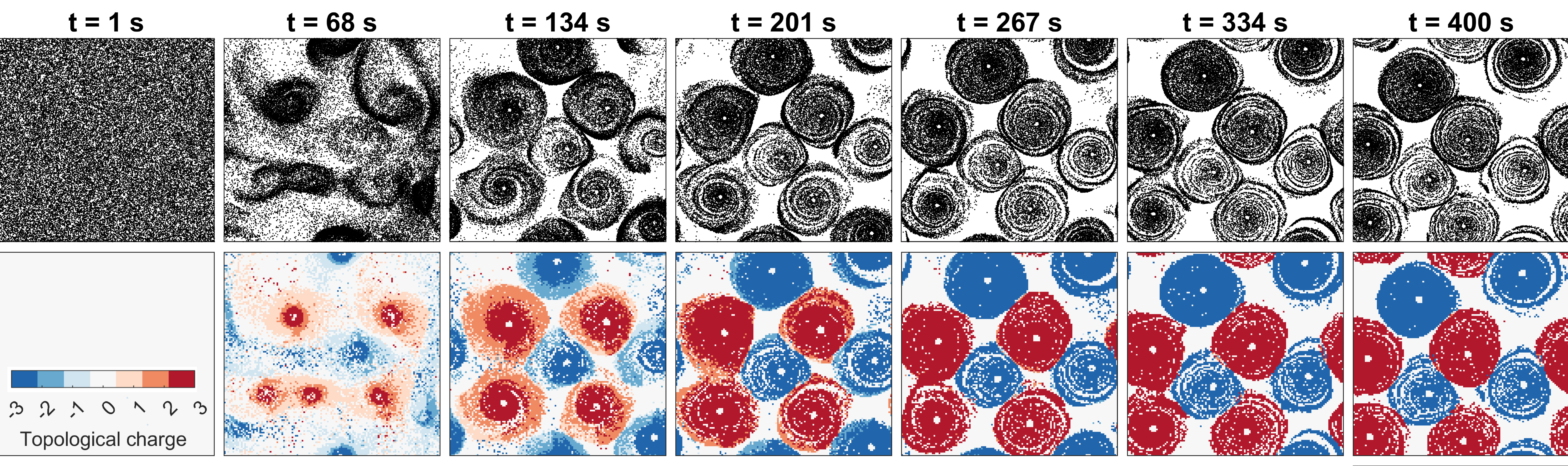}
    \caption{\textbf{Active vortex glass} made of particles whose raw distribution features an inverse cascade ($\Pi(k)<0$), but whose RFE field features a forward-cascade ($\Pi(k)>0$). The defect count falls off drastically and flattens; defect merger is arrested and the ensemble settles in a frozen defect lattice. Top row shows the RFE particle point-cloud while bottom row shows topological charge, while the columns correspond to seven temporal snapshots.}
    \label{fig:glass}
\end{figure*}

Finally, we observe that the formation of the macroscopic dipole serves as a robust dynamic attractor for the system. Figure~\ref{fig:1}d) maps the system's trajectory in the energy-order phase plane ($E$ vs. $P$). The system evolves away from the disordered origin (a random gas), increasing both its macroscopic kinetic energy and its dipole order parameter $P$, with the observed fluctuations reflecting the dynamic shock-merger mechanism. Viewed through the lens of the effective Hamiltonian (Eq.~\ref{eq:hamiltonian}), this trajectory structurally mimics an ascent up the entropy curve toward a negative-temperature-like regime \cite{onsager_statistical_1949}. Driven by the intrinsic disorder ($\omega_i$) of the active bath, the continuous injection of microscopic enstrophy pushes the system deeper into the tilted washboard potential identified in the Adler formalism \cite{ivarsenPolarChiralActive2026}. In the bounded domain, this sustained energy injection favors the continuous spectral condensation of vorticity, maintaining the macroscopic dipole as a stable dynamic steady-state rather than a strict thermodynamic equilibrium. The negative spectral flux in Figure~\ref{fig:1}b) supports the interpretation that the active particles act as a dissipative kinetic layer, driving an effective macroscopic fluid that exhibits transport dynamics analogous to those described by the conservative Kirchhoff-Onsager Hamiltonian. This framework establishes a theoretical bridge between the highly dissipative microphysics of chiral active matter and the conservative, large-scale transport phenomena characteristic of classical inviscid fluids.

\section{Discussion}

Our results offer a phenomenological framework that may help clarify the non-universal spectral scaling frequently observed in active matter \cite{james_turbulence_2018,linkmann_phase_2019,bourgoin_kolmogorovian_2020,mukherjee_intermittency_2023}. By investigating an aggressively abstracted model of polar chiral active matter, we observe a highly interpretable, scale-dependent duality. While some established literature associates steep dissipative spectra with large-scale coherence and shallower scaling with small-scale fluctuations \cite{chertkov_dynamics_2007,zhuFlowPatternsEnergy2024}, our phase-slaved model demonstrates a distinct, inverted dynamic. At microscopic scales ($k > k_c$), the agent ensemble is dominated by the tight packing of phase singularities, yielding a steep spectral slope around $-8/3$ \cite{kraichnan_inertial-range_1971,borue_spectral_1993,borue_inverse_1994}. Within the specific context of our model's Kuramoto-Sakaguchi-like interaction, we interpret this steepness in terms of enstrophy injection -- the continuous conversion of intrinsic chemical energy (represented by frustration $\omega_i$) into topological charge. By isolating the macroscopic transport from this active microscopic bath, we demonstrate how an effective inertial cascade can emerge, providing a potential analytical lens for future studies of natural systems grappling with complex, non-universal cascades.

\begin{table*}
\centering
\caption{\textbf{Thermodynamic Phases of Chiral Active Matter.} The system state is determined by the intrinsic frequency dispersion $\Delta\omega$. High activity drives the inverse cascade towards the inviscid dynamics, while low activity leads to kinetic jamming (glass). $\Pi_{Raw}$ and $\Pi_{RFE}$ refer to spectral flux for the raw particle distribution and the RFE field respectively. See Figure~\ref{fig:ensemble} for a characterization of the three phases; the robustness of the phases against the kernel size $\sigma$ is examined in Appendix~\ref{app:robustness}.
}
\label{tab:phase_diagram}
\begin{tabular*}{\textwidth}{@{\extracolsep{\fill}}l c l l l}
\toprule
\textbf{Phase} & \textbf{$\Delta \omega$} & \textbf{Micro-Physics} & \textbf{Macro-State} & \textbf{Signature} \\
\midrule
\textbf{I. Global synchronisation} & $\Delta \omega \to 0$ & $\nabla \phi \to 0$. & Static clumps (Fig.~\ref{fig:clumps}). & $\Pi_{Raw} > 0,\; \Pi_{RFE} > 0$ \\
\textbf{II. Active Vortex Glass} & Low / narrow & Frustration; no merger. & Defect solid  (Fig.~\ref{fig:glass}). & $\Pi_{Raw} < 0,\; \Pi_{RFE} > 0$ \\
\textbf{III. Onsager Condensate} & High / broad & Shock merger. & ``Onsager dipole''  (Fig.~\ref{fig:0}). & $\Pi_{Raw} < 0,\; \Pi_{RFE} < 0$  \\
\bottomrule
\end{tabular*}
\end{table*}


The macroscopic trajectory (Figure~\ref{fig:1}d) substantiates the foregoing. Driven by continuous enstrophy injection at the microscale, the renormalised fluid demonstrably climbs the energy-order phase plane. While this  clustering structurally mimics the entropy-maximizing transition to a negative-temperature regime characteristic of classical Onsager point-vortices \cite{onsager_statistical_1949}, in our driven-dissipative system it emerges as a dynamic, steady-state attractor rather than a strict thermodynamic equilibrium.

Crucially, the RFE framework provides a potential analytical tool for exploring the diverse, non-universal spectra frequently reported in biological and active matter experiments. Rather than viewing the sharp deviations from Kolmogorov scaling observed in bacterial swarms, active nematics, and cytoskeletal gels  strictly as a breakdown of fluidic transport, we propose that the raw particle fields may sometimes obscure coexisting macroscopic modes. While the specific spectral exponents of a given system will inevitably depend on its effective activity levels and underlying nonlinear dynamics (as demonstrated by Ref.~\cite{mukherjee_intermittency_2023}), analysing these systems via renormalised hydrodynamic modes might help decouple conservative macroscopic transport from microscopic chaos. By applying operators analogous to the RFE (Eqs.~\ref{eq:gc}, \ref{eq:gc2})-- essentially a low-pass filter -- future experimental and theoretical studies should test whether hidden inertial-like transport mechanisms are recoverable in a broader class of overdamped systems.  Eq.~(\ref{eq:RFEtheory}) in Appendix~\ref{app:hydro} provides a clear physical reasoning for this step.

\subsection{The role of intrinsic frustration}

The core dynamic of our model is dictated by $\omega_i$, or, in practical terms, the distribution $\Delta \omega$ shown in Figure~\ref{fig:distro}. Examining the role of this parameter allows us to map out the area of validity of the foregoing discussions.

The realization of the inverse cascade, and subsequent conservative dynamics, strictly depends on the system's ability to overcome topological energy barriers \cite{marov_self-organization_2013}. As in Ref.~\cite{ivarsenPolarChiralActive2026}, we find that the intrinsic frequency dispersion $\Delta \omega$ determines whether the agent ensemble freezes, jams, or flows, reflecting the depinning transition of the tilted washboard potential experienced by each agent, where $\Delta \omega$ acts as the effective bias current driving the system across the Adler-Ohmic bifurcation \cite{ivarsenPolarChiralActive2026}.

Figure~\ref{fig:glass} illustrates the foregoing by presenting the evolution of a simulation that used a narrower, lower-frequency dispersion $\Delta \omega$ (see Figure~\ref{fig:distro}); the raw particle distribution subsequently drove an inverse cascade, but the renormalised fluid did not, and the active bath was unable to transfer energy to wavenumbers $k<k_c$, resulting in an \textit{arrested topological defect glass state}. While the Onsager fluid exhibits continuous, power-law coarsening, the vortex glass state exhibits a rapid localized quench ($N\propto t^{-2}$) followed by a kinetic \textit{arrest} ($N\rightarrow \text{const.}$), and static, quantized loop currents of phase information \cite{delabays_multistability_2016}. This leads us to suggest that the experimental ``active glass'' described in Ref.~\cite{chardac_emergence_2021} may be attributable to the frustration, or \textit{noise}, associated with the system's constituents.

In Table~\ref{tab:phase_diagram} we summarize the results of systematically varying the shape of $\Delta\omega$. This yields the stability boundaries of our results, and, in what follows, we provide the empirical and theoretical foundation for the phase diagram in Table~\ref{tab:phase_diagram}. We shall do this through simulations that systematically vary the disorder distribution $\Delta \omega$. The complementary robustness test, varying the kernel size $\sigma$ at fixed $P(\omega)$, is presented in Appendix~\ref{app:robustness}.

\subsubsection{Three thermodynamic phases} \label{sec:tridiscussion}

Consider the theoretical turbulence frameworks of Ref.~\cite{marov_self-organization_2013} (self-organisation) and Ref.~\cite{chertkov_dynamics_2007} (forced hydrodynamics). The former describes an active, excitable medium of interacting fluctuations capable of self-organization, wherein a complex mean field amplitude ($Z$) that arises from phase synchronisation acts as an effective force that traps individual stochastic oscillators into coherent clusters. The latter modeled a stable, self-similar dipole condensate that accumulates inverse-cascading energy at the system size, yielding a large-scale coherent vorticity $\Omega$. 

\begin{figure}
    \centering
    \includegraphics[width=0.46\textwidth]{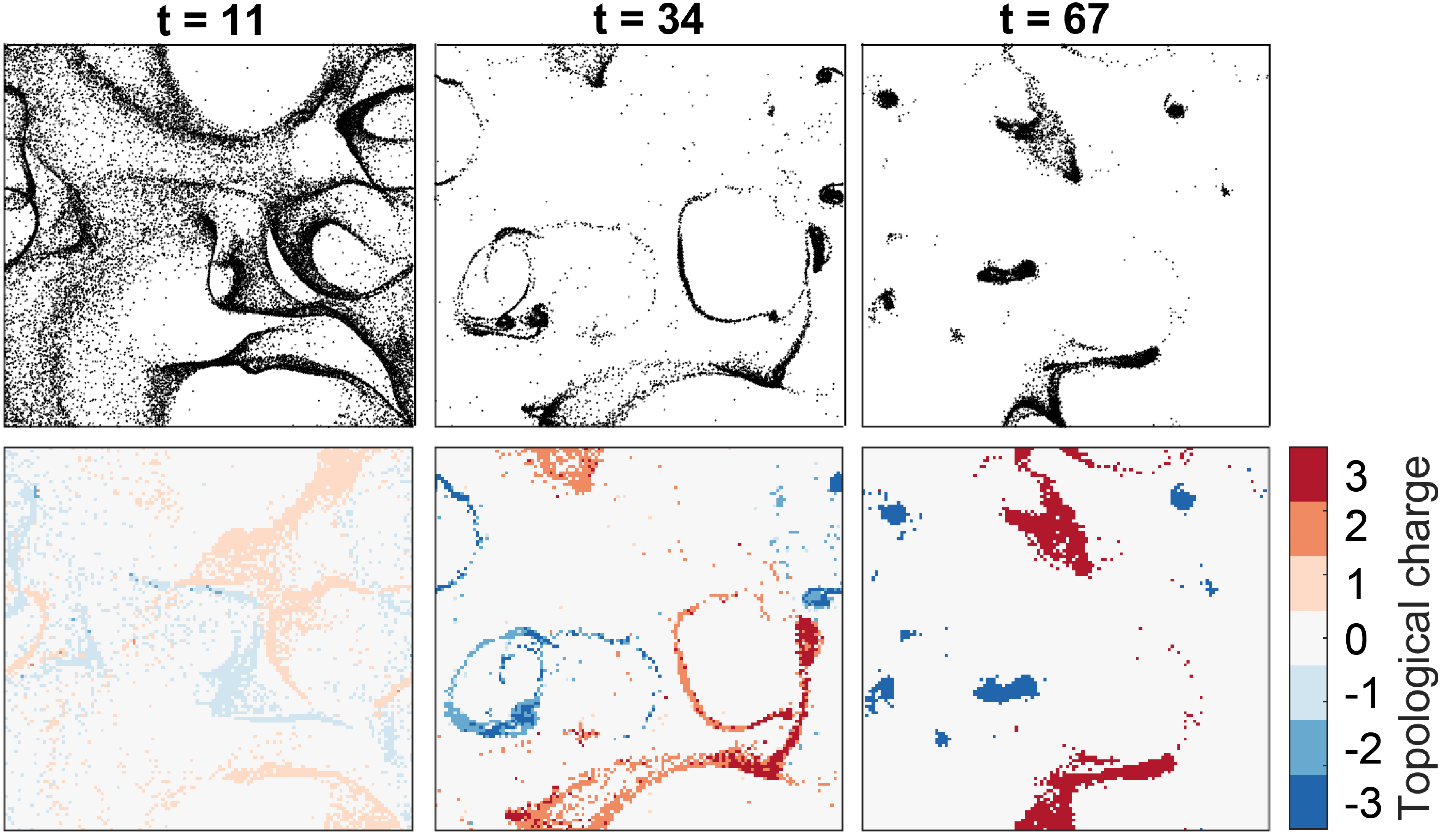}
    \caption{\textbf{Summary of a Phase I simulation.} RFE particles as point-clouds (top row) and topological charge (bottom row) mapped with a colorscale; timestamps are indicated.}
    \label{fig:clumps}
\end{figure}

\begin{figure*}
    \centering
    \includegraphics[width=\textwidth]{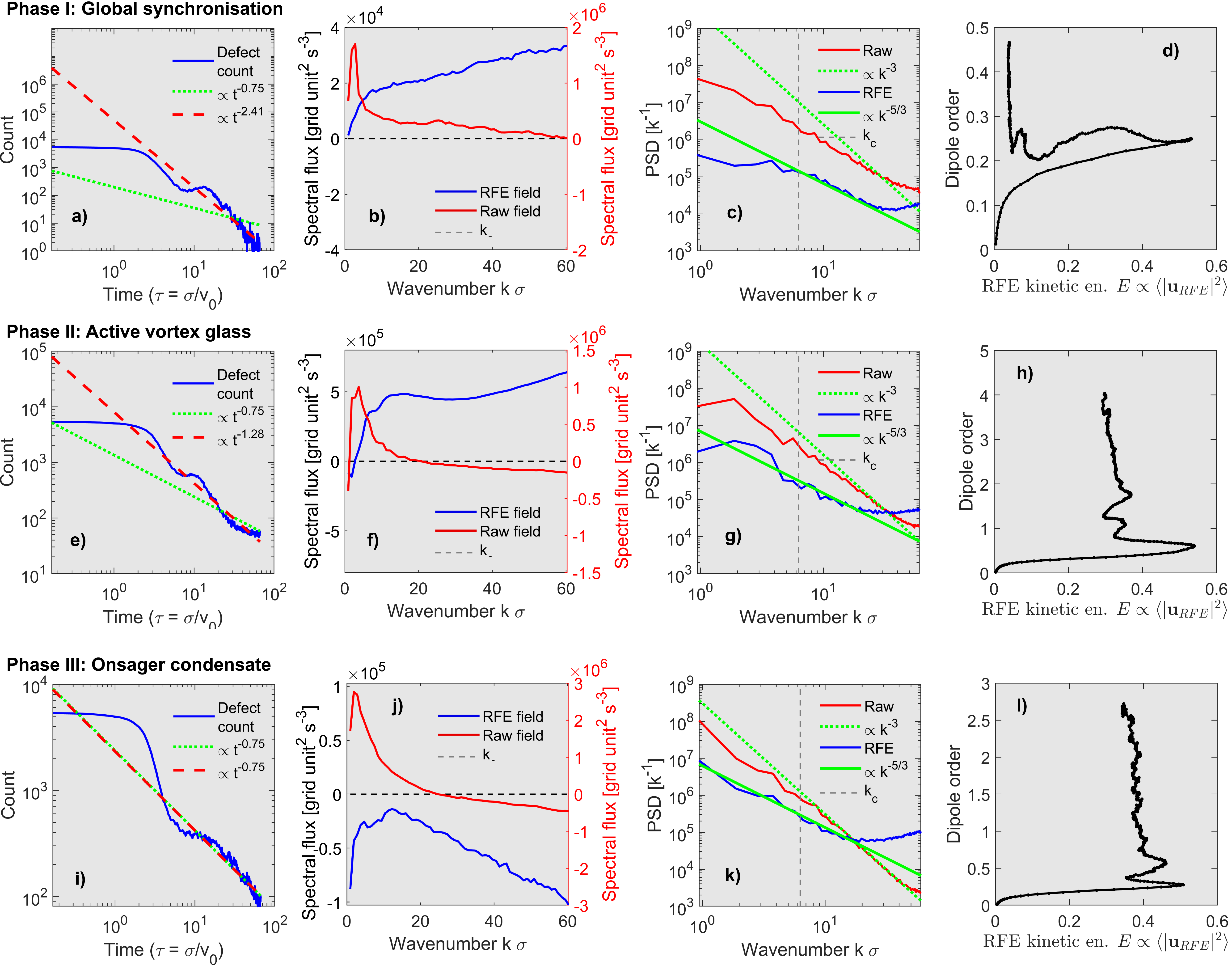}
    \caption{\textbf{Ensemble averages of 16 simulations, showing the behaviour of the three phases in Table~\ref{tab:phase_diagram}.} The leftmost column (panels a, e, and i) show the defect count $N_d(t)$ in a solid blue line, with a $t^{-0.75}$ powerlaw in dotted green line and a fit for $t>3.4$ (through non-linear minimization of root-mean square error) powerlaw in dashed red line. The next column (panels b, f, and j) show median spectral flux for the raw (red) and RFE (blue) fields. The third column (panels c, g, and k) show the energy spectrum for the raw (red) and RFE (blue) fields, with $k^{-3}$ and $k^{-5/3}$ powerlaw fits, as well as the cut-off scale $k_c$. The last column (panels d, h, and l) show the thermodynamic trajectory. All quantities are shown as medians of 16 randomized (in initial locations $x_i,y_i$ and white-noise $\eta_i$) runs. See Appendix~\ref{app:robustness} for the dependence of the raw-field spectral slope on scale separation. \label{fig:ensemble}}
\end{figure*}

Before we pinpoint the conjunction between the latter two theoretical forays, we must comment on the three distributions in $\omega_i$ on display in Figure~\ref{fig:distro}, corresponding to $\Delta\omega$ in Table~\ref{tab:phase_diagram}. Intrinsic frustration $\omega_i$ near the box-size, $(2\Lambda)^{-1}$, acts simultaneously as $Z$ in Ref.~\cite{marov_self-organization_2013} and $\Omega$ in Ref.~\cite{chertkov_dynamics_2007}. Thus, if frustration $\Delta \omega$ is too low-valued and too narrow, the system generates active stress but lacks the transport coherence to move it, yielding an arrested vortex glass. Conversely, if the dispersion in $\omega_i$ is both sufficiently high and sufficiently broad (reaching near-box-size fluctuations), the system effectively facilitates the emergence of a space-filling dynamic dipole, akin to an Onsager dipole. In Figure~\ref{fig:ensemble} we characterize the foregoing descriptions by extracting the median results from 16 ensemble simulations of the three phases. Crucially, we observe that the spectral flux in Phase I (see Figure~\ref{fig:clumps}) is positive for both the raw and RFE fields, while for Phase II (Figure~\ref{fig:ensemble}f), the flux is negative for the raw field (crossing zero at intermediate wavenumbers) and positive for the RFE field, and for Phase III the flux is consistently negative, facilitating the inverse cascade that would otherwise be arrested in the case of Phase II.

Thus, the inviscid dynamics we observe allow the system to access the condensate state. This suggests that Ref.~\cite{marov_self-organization_2013}'s thermodynamic driver is strong enough in this regime to overcome the chaotic transients \cite{giomi_defect_2013} and access Ref.~\cite{chertkov_dynamics_2007}'s equilibrium basin. This is in contrast with the active nematic systems that often saturate in a state of defect chaos \cite{giomi_defect_2013}. The conjunction of Ref.~\cite{marov_self-organization_2013}’s extremum principles with Ref.~\cite{chertkov_dynamics_2007}’s statistical mechanics bolsters the notion that the inviscid dynamics in our simulation is a thermodynamically preferred attractor.

\subsubsection{Constraints on the driver distribution} \label{sec:constraints}

Two questions naturally arise regarding $P(\omega)$: what constraints, beyond the height and broadness criteria identified above, must the distribution satisfy, and in what sense is the resulting fluid chiral? On the first question, the power-law form adopted here is studied systematically in Ref.~\cite{ivarsenPolarChiralActive2026} through the index $n$; within the present study we have identified no further necessary constraints on the shape of $P(\omega)$, provided the dispersion reaches near-box-size fluctuation frequencies (Figure~\ref{fig:distro}); the scale-free form is what the height and broadness criteria implicitly require. In this sense, $\Delta\omega$ constitutes an effective temperature for the ensemble \cite{ivarsenPolarChiralActive2026}. We have not, however, systematically varied the sign symmetry of the distribution.
On the second question, our driver distribution is symmetric about zero, and so the ensemble contains statistically equal populations of clockwise and counterclockwise rotors, and the global mirror symmetry of the system is unbroken in the ensemble-averaged sense. Chirality in our model is therefore an agent-level property, and macroscopic parity breaking emerges locally, through the spontaneous segregation of like-signed $\omega_i$ into distinct bands and domains (Figure~\ref{fig:topography}a, b); the Onsager dipole is the system-scale limit of this segregation. A closely comparable phenomenology arises in spinner fluids composed of mixed clockwise and counterclockwise populations, which segregate into lanes in the inertialess limit and exhibit turbulent-like motion with $k^{-5/3}$ scaling at small but finite Reynolds number \cite{reevesEmergenceLanesTurbulentlike2021}. A strictly one-signed driver would inject net angular momentum, and, by analogy with the chirality-breaking transition reported in Ref.~\cite{reevesEmergenceLanesTurbulentlike2021}, we expect it to bias the condensate toward a single-signed rotating state rather than the dipole reported here; we leave this as an open question.

\subsection{Marginal synchronisation in nature}

The hydrodynamic limit in our simulations is thus in a \textit{marginal synchronisation regime}, meaning that the synchronising force is balanced by continuous, shock-driven thermalisation of inertial energy. This is empirically confirmed by observing that the dynamics in Figure~\ref{fig:0} effectively \textit{never stop}. The observed density decay $N(t) \sim t^{-0.75}$ during this phase favors coherent structure merger in shallow waters \cite{larichev_weakly_1991} over compressible gas dynamics: the extended runs in Appendix~\ref{app:robustness} exclude the $t^{-2/3}$ and $t^{-1}$ rates \cite{kida_asymptotic_1979} at the 95\% level. The dominance of shock-mergers over smooth thinning indicates that the system operates in the supersonic regime ($M>1$), which acts as a direct analog to supercritical shallow water flow, where density fluctuations map directly to hydraulic jumps \cite{unruh_experimental_1981, hoefer_dispersive_2006}.

Future studies should map out the biological, chemical, and environmental circumstances that facilitate the emergence of an inertial flow in soft, active matter turbulence, and when such systems are dominated by low-Reynolds number friction and rapid energy decay \cite{toner_flocks_1998,wensink_emergent_2012,giomi_defect_2013}, as well as specific instances where the theoretical foundations for our hydrodynamic limit breaks down.


In the biological microscopic realm, large-scale mixing is vital for nutrient transport \cite{jeanneret_entrainment_2016}. Conceptually, our model suggests that a biological system, for reasons external or internal, benefits from a specific variance in intrinsic traits (modeled as the dispersion $\Delta \omega$), a heterogeneity needed to activate the topological heat pump, or risk reverting to a lattice of topological defects. This may contribute to an explanation for why phenotypic heterogeneity is prevalent \cite{ariel_swarming_2015} and sufficiently extreme, and suggests that such noise expressions are in fact beneficial for e.g., cohesion or efficient nutrient circulation, and other traits associated with flocking. Biological systems may thus inadvertently exploit noise in gene expression \cite{ackermann_functional_2015}, effectively tapping into disorder as a means for macroscopic transport (see, e.g., Ref.~\cite{marov_self-organization_2013}).

In our model, when the system is disordered ($R\to0$), it is effectively massless and governed by overdamped diffusion. As it synchronises, it acquires mass ($\lambda = R^2$), supporting momentum transport and waves akin to correlation propagation in flocks and swarms \cite{toner_flocks_1998,brumley_flagellar_2014}, with the caveat that our system operates in two dimensions. The observed $R$-dependency implies that defects behave as \textit{massless holes} (see Figure~\ref{fig:topography} in Appendix~\ref{app:hydro}) in an otherwise massive fluid. This mass contrast enables supersonic defect kinematics: phase singularities accelerate across phase gradients faster than the bulk sound speed \cite{attanasi_information_2014}, maintaining scale-free coherence \cite{cavagna_scale-free_2010} without central processing, in flocks and other instances of dry chiral active matter.

\section*{Acknowledgements}
This work is supported by the European Space Agency’s Living Planet Grant No. 1000012348. The author is grateful to O. Nestande, D. Knudsen, PT. Jayachandran, and K. Douch for stimulating discussions. \\

Google's Gemini 3.0 Pro and Anthropic's Claude Fable 5 have been used to assist mathematical formalism and coding in \textsc{matlab}.\\

A collection of videos is available in the Supplementary Materials online.

\section*{Data Availability}

The code that supports the findings in this paper is openly available at Zenodo \cite{ivarsenDryPolarChiral2026}.

\appendix

\section{Robustness of the phase diagram} \label{app:robustness}

This appendix presents some additional numerical tests of the inferences made in the Discussion section, and in relation to Table~\ref{tab:phase_diagram} and Figure~\ref{fig:ensemble}. First, we show that the vortex thinning observed in our simulations, and the theorized shock merger, which depends on the $-0.75$~exponent, is largely consistent. Through an ensemble of extended simulations (from 400~s to 2000~s simulation time), we confirm the consistency and demonstrate a tendency for an exponent of $-3/2$ or $-2$ to be excluded.

After this, we present additional simulations performed by changing the kernel size $\sigma$, keeping all other factors unchanged, elucidating the role ascribed intrinsic noise, or heterogeneity, on the various stable phases in the model dynamics.

\subsection{Extended defect-count decay}

\begin{figure}
    \centering
    \includegraphics[width=0.5\textwidth]{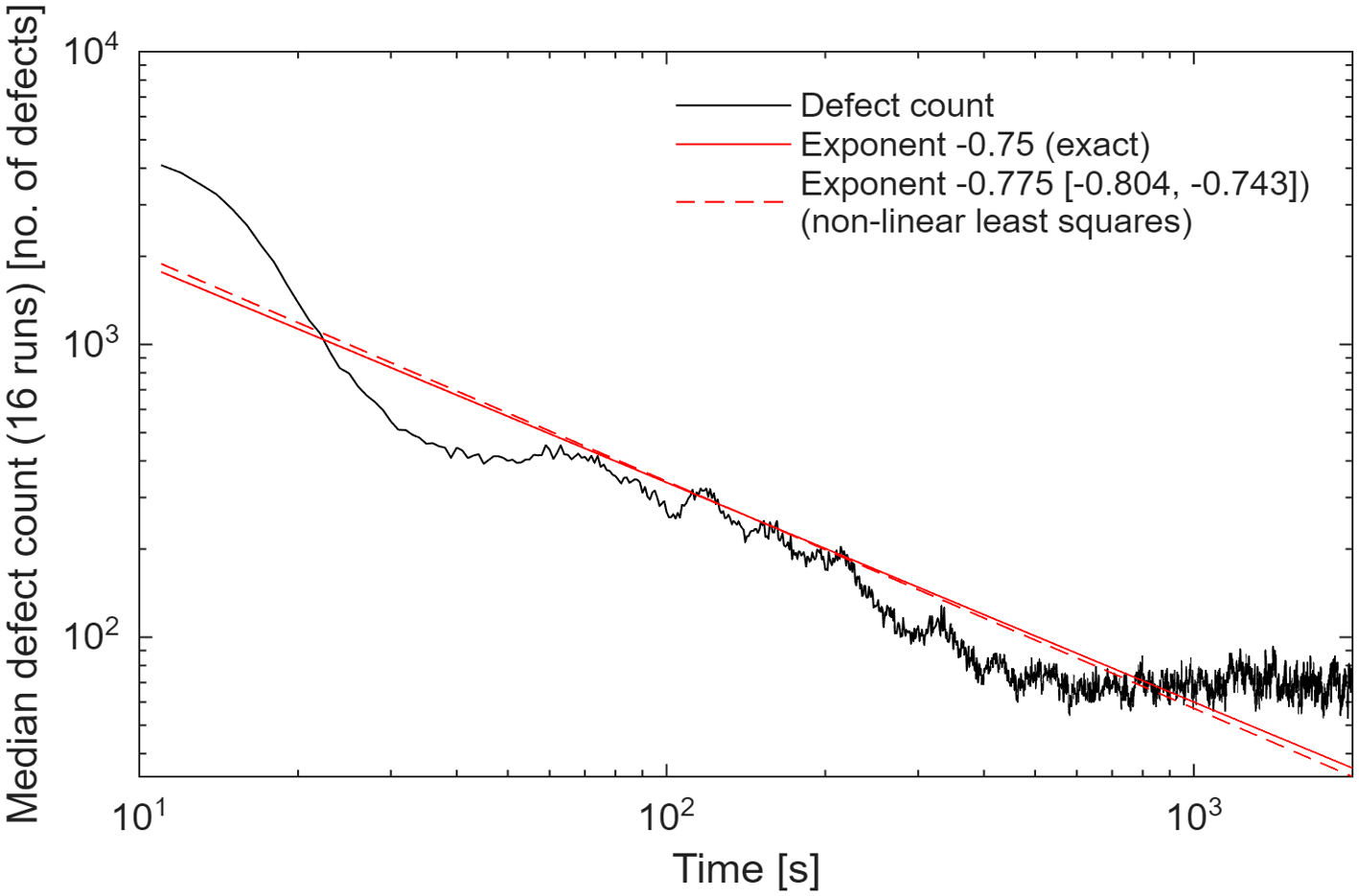}
    \caption{\textbf{Median defect count of 16 extended Phase~III runs} ($t \le 2000$~s), shown for $t > 10$~s. The solid red line shows the exact $t^{-0.75}$ scaling; the dashed red line shows the unconstrained fit, $t^{-0.776}$, with bootstrapped 95\% confidence interval $[-0.804, -0.743]$. The fit window is the full plotted range.}
    \label{fig:decay}
\end{figure}

To address the short scaling range of the defect-count decay in Figures~\ref{fig:1}a and \ref{fig:ensemble}, we extended the Phase~III ensemble to 16 runs evolved to $t = 2000$~s. Figure~\ref{fig:decay} shows the median defect count for $t > 10$~s. An unconstrained log-log least-squares fit over the full plotted range yields an exponent of $-0.776$; a bootstrap over the 16 ensemble members gives a 95\% confidence interval of $[-0.804, -0.743]$. The interval contains the theoretical $-3/4$ \cite{carnevale_evolution_1991,larichev_weakly_1991} and excludes both the diffusive $t^{-1}$ rate and the compressible $t^{-2/3}$ rate \cite{kida_asymptotic_1979}. Beyond $t \approx 800$~s the count flattens at $N_d \approx 70$; we identify this floor with the steady-state balance between noise-driven pair creation and shock merger that characterizes the marginally synchronised attractor of the main text: the decay does not proceed to zero, consistent with the dynamics never ending.

\subsection{Kernel-size sweep}

\begin{figure*}
    \centering
    \includegraphics[width=0.88\textwidth]{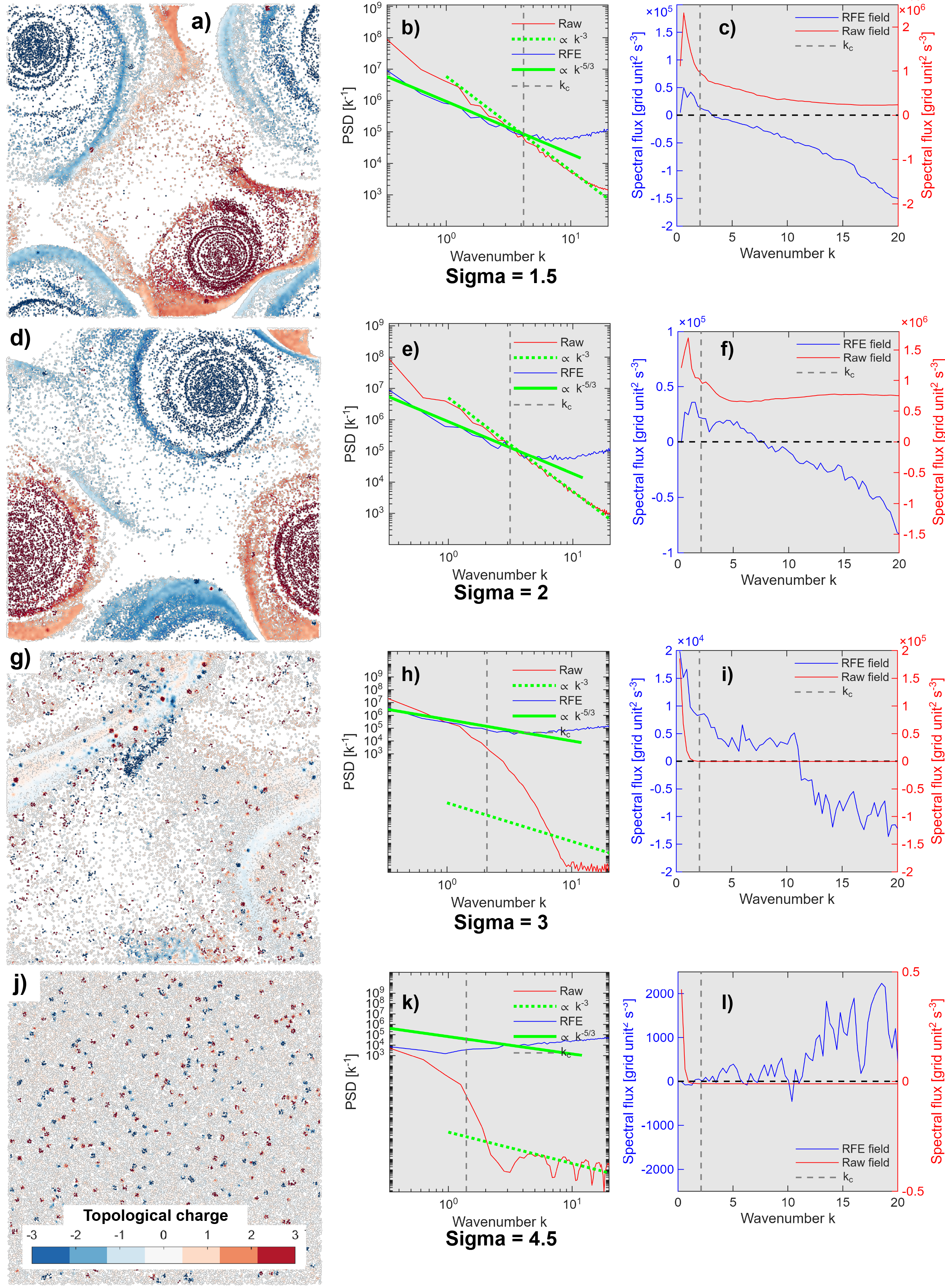}
    \caption{\textbf{Kernel-size sweep at fixed $P(\omega)$}, for $\sigma = 1.5, 2, 3, 4.5$ (columns; $\Lambda \approx 27, 20, 13, 9$). Rows show end-state oscillator point-clouds, energy spectra for the raw (red) and RFE (blue) fields with $k^{-3}$ and $k^{-5/3}$ guide lines and $k_c = 2\pi/\sigma$ (dashed grey), median spectral flux for both fields.
    }
    \label{fig:sweep}
\end{figure*}

To test whether the phenomenology of the main text is contingent on the interaction kernel, we repeated the Phase~III simulation with $\sigma$ varied over $\{1.5, 2, 3, 4.5\}$ while holding $P(\omega)$ fixed in physical units. The RFE averaging time $\tau = \sigma/v_0$ and the cutoff $k_c = 2\pi/\sigma$ co-vary with $\sigma$ accordingly, and the kernel support is scaled with $\sigma$; all other parameters are as in the main text. Figure~\ref{fig:sweep} summarizes the results.

We make the following observations. \textbf{\textit{(1)}} Coherent vortex clustering and a negative RFE spectral flux persist for $\sigma = 1.5$--$3$. The morphology of the condensate depends on the scale separation $\Lambda = 2L/\sigma$: several distinct vortices form at $\Lambda \approx 27$ and $20$, while the single space-filling dipole of the main text obtained at $\Lambda \approx 13$. \textbf{\textit{(2)}} At $\sigma = 4.5$ ($\Lambda \approx 9$), the defect population collapses shortly after initialization, and no large-scale structuring forms; the RFE flux is here positive, and the reported phenomenology not triggered thus requires sufficient scale separation. \textbf{\textit{(3)}} The raw-field spectrum steepens from approximately $k^{-8/3}$ at $\sigma = 3$ toward $k^{-3}$, sustained over roughly a decade, at $\sigma = 1.5$ and $2$; the steeper $-8/3$ value quoted in the main text thus appears as a finite-scale-separation result. \textbf{\textit{(4)}} The RFE spectrum remains consistent with $k^{-5/3}$ below $k_c$ wherever the condensate forms, and the sub-$k_c$ range widens as $k_c$ is raised. \textbf{\textit{(5)}} The magnitude of the spectral flux decreases with increasing $\sigma$. We conclude that the phase behaviour of Table~\ref{tab:phase_diagram} is not set by $\sigma$; the kernel size sets the scale separation $\Lambda$, and thereby the number and size of the coherent structures, consistent with the phases being controlled by the disorder distribution $\Delta\omega$ \cite{ivarsenPolarChiralActive2026}.

\section{Derivation of the hydrodynamic limit} \label{app:hydro}

In this appendix, we derive the hydrodynamic limit of our model, beginning with deriving the macroscopic velocity, the advective transport term for the model, and establishing an analogy with shallow water hydrodynamics, providing closure to the argument that active chiral matter can exhibit Euler turbulence. We note that we reinstate the units in this section to track dimensional consistency.

\subsection{Velocity and continuity}

We define the macroscopic velocity field $\mathbf{u}(\mathbf{r},t)$ as the local average of the particle velocities. Decomposing the particle phase into the local mean phase $\Psi$ and a fluctuation $\delta \phi_i$ (that is, $\phi_i = \Psi + \delta \phi_i$), and assuming symmetric fluctuations (that is, $\langle \sin \delta \phi \rangle \approx 0$), the coarse-grained velocity becomes,
\begin{equation} \label{eq:RFEtheory}
    \mathbf{u} = \langle \mathbf{v}_i \rangle \approx v_0 \langle \cos \delta \phi \rangle (\cos \Psi, \sin \Psi).
\end{equation}
We identify the local order parameter $R_1$,
\begin{equation}
    R_n = \langle \cos(n \delta \phi) \rangle = \text{Re}\left[ \langle e^{i n \delta \phi} \rangle \right],
\end{equation}
for a peaked (narrow) Gaussian distribution in $\delta\phi$; we then drop the subscript. $R$ measures the degree of local synchronisation. Thus, we obtain,
\begin{equation} \label{eq:macro}
    |\mathbf{u}| = v_0 R.
\end{equation}
While individual particles move at constant speed $v_0$, the renormalised fluid element speed is variable and proportional to the local synchronisation. To express $\mathbf{u}$ in terms of the mean field  $\Psi$, we identify the flow direction $\mathbf{\hat{n}} = (\cos \Psi, \sin \Psi)$ with the normalized phase gradient:
\begin{equation} \label{eq:u}
    \mathbf{u} = v_0 R \frac{\nabla \Psi}{|\nabla \Psi|}.
\end{equation}
Eq.~(\ref{eq:u}) is globally valid: at defect cores where the phase $\Psi$ is undefined, the order parameter vanishes ($R \to 0$), ensuring the velocity field remains continuous and bounded ($\mathbf{u} \to 0$). Recognizing that $\nabla \Psi$ scales as $1/r$ near defects, we note that the vector $\mathbf{u}$ is parallel to $\nabla \Psi$. Far from defect cores, where the phase gradient magnitude varies slowly ($|\nabla \Psi| \approx \text{const}$), the flow simplifies to an effective potential flow. By identifying the effective circulation quantum $\kappa=v_0\sigma$ (see Eq.~\ref{eq:kappaid}), demanded by the phase stiffness of the agent description as a motile, disordered Josephson junction array \cite{ivarsenPolarChiralActive2026}, we write:
\begin{equation} \label{eq:flow}
    \mathbf{u} \approx \kappa R \nabla \Psi = v_0\sigma R \nabla \Psi.
\end{equation}
This approximation eventually allows us to utilize the machinery of inviscid fluid mechanics in the bulk of the domain, while the factor $R$ naturally regularizes the dynamics at the topological singularities.

Before providing closure to the above, we need to address continuity. The continuity equation for $\mathbf{u}$ then reads,
\begin{equation} \label{eq:continuity}
    \frac{\partial \rho}{\partial t} + \nabla \cdot (\rho \mathbf{u}) = 0.
\end{equation}
Substituting Eq.~(\ref{eq:flow}) into Eq.~(\ref{eq:continuity}), we observe that the macroscopic density evolution is governed by mean phase curvature:
\begin{equation} \label{eq:curvature_transport}
    \frac{\partial \rho}{\partial t} \approx - v_0 R \nabla \cdot (\rho \nabla \Psi).
\end{equation}
This formulation identifies the system as formally analogous to the shallow water equations \cite{tan_shallow_1992,vreugdenhil_numerical_2013}, where the active density $\rho$ acts as the fluid height (see Sec.~\ref{sec:shallow} in Appendix~\ref{app:hydro}). The phase curvature $\nabla^2 \Psi$ drives the compression of the fluid elements, explicitly linking topological defects to density fluctuations. 

Eq.~(\ref{eq:winding}) implies $|\nabla \Psi| = m/r$. Substituting this into our macroscopic velocity definition (Eq. \ref{eq:u}), and assuming the order parameter $R$ is \textit{locally} uniform, the fluid velocity induced by a defect scales as:
\begin{equation} \label{eq:rnumerator2}
    |\mathbf{u}| \approx v_0 R \frac{m}{r}.
\end{equation}
This confirms that the topological defects of the mean phase act as point vortices with an effective physical circulation $\Gamma \approx 2\pi m v_0 R$. Consequently, we arrive at the Kirchhoff-Onsager Hamiltonian in Eq.~(\ref{eq:hamiltonian}).  
Crucially, this derivation highlights that the topological charge driving the Hamiltonian dynamics is scaled by the order parameter $R$. If renormalization (or synchrony) is suppressed ($R \to 0$), the effective circulation vanishes ($\Gamma_{\text{eff}} \to 0$), causing the Hamiltonian interaction to collapse.

\subsection{Momentum and pressure}

We start with the discrete, slaved velocity,
\begin{equation*}
    \mathbf{v}_i = v_0 (\cos \phi_i, \sin \phi_i).
\end{equation*}
We define a momentum density field $\mathbf{g}(\mathbf{r},t)$, which in the discrete case becomes, 
\begin{equation}
    \mathbf{g}(\mathbf{r},t) = \sum_{i=1}^N \mathbf{v}_i(t) \delta(\mathbf{r} - \mathbf{r}_i(t)),
\end{equation}
where $\delta(x)$ now defines the Dirac delta function. Next, take the time derivative $\partial_t \mathbf{g}$,
\begin{equation} \label{eq:2}
    \frac{\partial \mathbf{g}}{\partial t} = \underbrace{\sum_i \dot{\mathbf{v}}_i \delta(\mathbf{r}-\mathbf{r}_i)}_{\text{Orientation}} + \underbrace{\sum_i \mathbf{v}_i \frac{\partial}{\partial t}\delta(\mathbf{r}-\mathbf{r}_i)}_{\text{Transport}\equiv \mathcal{T}},
\end{equation}
where we identify the orientation term by noting that $\dot{\mathbf{v}}_i$ contains the Kuramoto-like phase dynamics (Eq.~\ref{eq:kuramoto}). The second term is identified as momentum transport by the chain-rule property of the delta function, $\partial_t \delta(\mathbf{r}-\mathbf{r}_i) = -\nabla \cdot (\dot{\mathbf{r}}_i \delta)$, as well as noting that $\dot{\mathbf{r}}_i = \mathbf{v}_i$,
\begin{equation}
    \mathcal{T} = - \nabla \cdot \left( \sum_i \mathbf{v}_i \mathbf{v}_i \delta(\mathbf{r}-\mathbf{r}_i) \right).
\end{equation}
Simplifying the notation, we write,
\begin{equation} \label{eq:transport0}
    \mathcal{T} = -\nabla \cdot \langle \mathbf{v}_i \mathbf{v}_i \rangle.
\end{equation}

To see how Eq.~(\ref{eq:transport0}) gives us an advective term $(\mathbf{u} \cdot \nabla) \mathbf{u}$,  we must calculate the tensor $\langle \mathbf{v}\mathbf{v} \rangle$. Consider the outer product $\mathbf{v}_k \mathbf{v}_k$ and average it; specifically,  the components (e.g., $xx$):
\begin{equation}
    \langle v_x v_x \rangle = v_0^2 \langle \cos^2(\Psi + \delta\phi) \rangle,
\end{equation}
using the expansion of $\phi_i$ as a perturbation to the mean field. Using trigonometric identities and averaging over the fluctuations, we get,
\begin{equation} \label{eq:vxvx}
    \langle v_x v_x \rangle \approx \frac{v_0^2}{2} [1 + \underbrace{\langle \cos(2\delta\phi) \rangle}_{R_2} \cos(2\Psi)],
\end{equation}
where we identify the second-order order parameter $R_2 \approx R^4$ (again assuming a peaked distribution), yielding an expression for \textit{microscopic} stress. In the fluid phase (the ``Onsager dipole'', Phase III), where the system exhibits marginal synchronisation ($0 < R < 1$), the fluctuations are dominated by frustration $\omega_i$, distinct from, and in opposition to, the synchronisation forces that maintain $R$.

For the equivalent expression for macroscopic stress, we use the definition of macroscopic velocity (Eq.~\ref{eq:macro}),
\begin{equation} 
    u_x u_x = (v_0 R \cos \Psi)^2 = v_0^2 R^2 \cos^2 \Psi.
\end{equation}
Trigonometry again allows us to isolate the anisotropic part,
\begin{equation} \label{eq:uxux}
    u_x u_x = \frac{v_0^2 R^2}{2} [1 + \cos(2\Psi)].
\end{equation}

To relate the foregoing to the hydrodynamic equations, we compare the microscopic and macroscopic tensors. In order to do so, we posit a constitutive relation of the form $\langle v_i v_j \rangle = P_{\text{eff}}\sigma_{ij} + \lambda u_i u_j$, where $\lambda$ scales the advective term. Substituting the expansions from Eqs.~(\ref{eq:vxvx})  and  (\ref{eq:uxux}) into this ansatz for the $xx$-component yields:
\begin{equation} \label{eq:posit}
    \underbrace{\frac{v_0^2}{2} + \frac{v_0^2 R^4}{2} \cos(2\Psi)}_{\text{Microscopic}} = \underbrace{P_{\text{eff}} + \lambda \left[ \frac{v_0^2 R^2}{2} + \frac{v_0^2 R^2}{2} \cos(2\Psi) \right]}_{\text{Macroscopic Closure}}.
\end{equation}
Matching the \textit{anisotropic} coefficients (the $\cos2\Psi$ terms) determines the coherence coefficient:
\begin{equation} \label{eq:inertia2}
    \frac{v_0^2 R^4}{2} = \lambda \frac{v_0^2 R^2}{2} \implies \lambda = R^2,
\end{equation}
confirming that the \textit{effective inertial mass} $\lambda$ scales with the square of the order parameter. 


While the foregoing provided a useful and instructional way to explicitly derive the emergent inertial mass through stress-tensor matching, a more rigorous derivation follows from the isomorphism to the Josephson junction array established in Ref.~\cite{ivarsenPolarChiralActive2026}. In this framework, the macroscopic superconductivity maintains a stiffness torque mediated by the complex order parameter field $Z = R e^{i\Psi}$. Crucially, Ref.~\cite{ivarsenPolarChiralActive2026} identified that the agent dynamics are formally isomorphic to the Adler equation, meaning individual agents move in a tilted washboard potential \cite{ambegaokar_voltage_1969},
\begin{equation} \label{eq:veff}
    V_\text{eff}(\phi_i) = -a_0R\cos(\Psi-\phi_i) - \omega_i\phi_i.
\end{equation}
According to the Ginzburg-Landau formalism for phase transitions, the free energy cost of sustaining a phase supercurrent (the superfluidic stiffness) scales with the squared magnitude of the complex order parameter field \cite{tinkham_introduction_2004},
\begin{equation} \label{eq:kappaid}
    \mathcal{E}_{kin} = \frac{1}{2} |Z|^2 (\nabla \Psi)^2 = \frac{1}{2} R^2 (\nabla \Psi)^2.
\end{equation}
We identify the gradient term as the superfluid velocity $v_s=\kappa\nabla\Psi$ (where $\kappa$ is a circulation quantum). By comparing this energy density to the hydrodynamic kinetic energy form $\lambda v_s^2/2$, we immediately recover the identity,
\begin{equation} \label{eq:inertia3}
\lambda \equiv R^2,
\end{equation}
confirming that the effective inertial mass $\lambda$ derived in our hydrodynamic limit (Eq.~\ref{eq:inertia2}) corresponds physically to the superfluid density (or phase stiffness) of the underlying Josephson junction array.

It is pertinent to compare this result to the swim pressure derived in the kinetic theory of active fluids (e.g., \cite{takatori_swim_2014}). 
In standard active fluids, the particle mass is constant (corresponding to $\lambda=1$), meaning the coherent kinetic energy scales as $u^2 \propto R^2$. In our topological gas, the effective inertial mass \textit{itself depends on local synchronisation} ($\lambda = R^2$), causing the ``coherent energy'' to scale as $\lambda u^2 \propto R^4$.


Next, we substitute the closure back into the momentum conservation equation and obtain,
\begin{equation} \label{eq:final}
    \frac{\partial \mathbf{u}}{\partial t} + \lambda \nabla \cdot (\mathbf{u} \mathbf{u}) = -\nabla P_{\text{eff}} + \mathbf{F}_{\text{active}},
\end{equation}
where we, for the time being, leave the exact nature of $\mathbf{F}_{\text{active}}$ ambiguous, as we must first establish the specific hydrodynamic analogy.


\subsection{Shallow waters and the active enthalpy functional} \label{sec:shallow}

\begin{figure}
    \centering
    \includegraphics[width=0.5\textwidth]{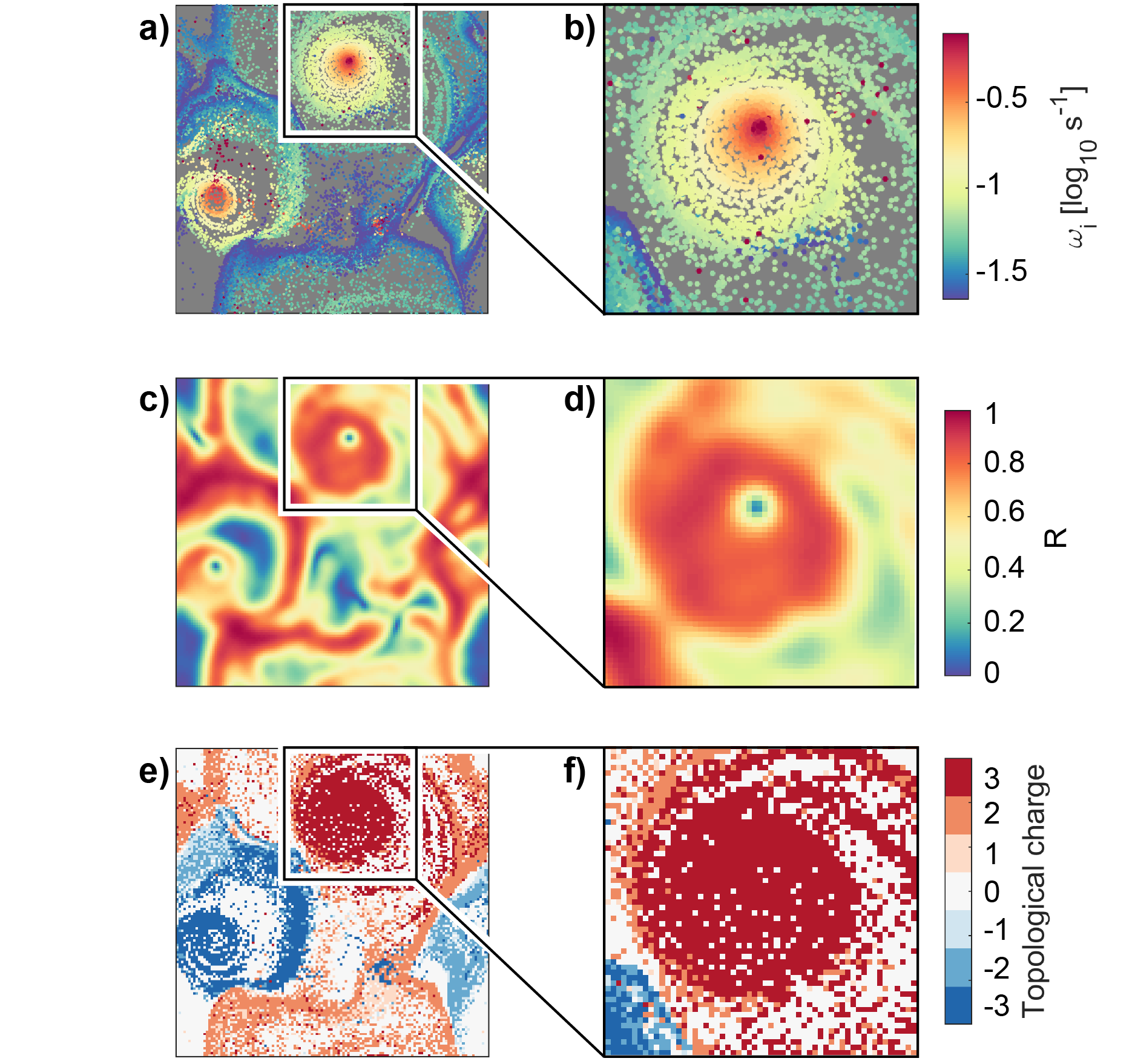}
    \caption{The end-state of a Phase~III (``Onsager dipole'') simulation, showing RFE particle locations, with colorscale denoting intrinsic frequency or frustration $\omega_i$ (panels a, b), the degree of local synchronization $R$ (panels c, d), and the topological charge (panels e, f). The right column shows a blown-up portion of the left column.}
    \label{fig:topography}
\end{figure}

As stated in the Methodology section, the interpretation of Eq.~(\ref{eq:curvature_transport}) identifies the system being modeled as conceptually and mathematically analogous to the shallow water framework, where the active density $\rho$ acts as the fluid height.
To proceed, we shall use the active enthalpy functional as a theoretical vehicle \cite{zhang_dynamic_2016,turzi_active_2017}. However, we must first bolster the connection with a specific empirical observation. Figure~\ref{fig:topography} shows a simulation of emergent inviscid dynamics, color-coding $\omega_i$ (top), $R$ (middle), and topological charge $m$ (bottom). We make two important observations;

\textit{(1)} As alluded to in the main text, $\omega_i$ in Figure~\ref{fig:topography}a, b) is effectively segregated into distinct bands and regions. Because of the segregation, the local average $\langle \omega_i \rangle$ at a position $\mathbf{r}$ becomes non-random and stable over timescales relevant to the flow. We can effectively ``smear'' the discrete particles into a continuous background field where $\omega(\mathbf{r}) \approx \langle \omega_i \rangle_{\mathbf{r}}$.

\textit{(2)} Figure~\ref{fig:topography}a--d) present evidence that, at topological defect cores, $R\to0$ and $\omega_i \gg \bar{\omega}$, which, along with Eq.~(\ref{eq:rnumerator2}), supports Eq.~(\ref{eq:u}). As we shall motivate below, the conditions at defect cores imply that the fluid is stiff. The flow must immediately adjust to the \textit{topography} defined by $\omega(\mathbf{r})$. The segregation of the active particles constitutes a potential $\Phi(\mathbf{r})$ that forces the flow.

Having motivated a conceptual shift from a Lagrangian to an Eulerian perspective, we define  the effective fluid density as $\rho(\mathbf{r}, t) \equiv R(\mathbf{r}, t)^2$. We then use $\omega(\mathbf{r})$ as an external scalar potential produced by intrinsic frequency mismatch $\omega_i-\bar{\omega}$, where $\bar{\omega}$ is the mean frequency. The ensemble average of this quantity then equals dispersion $\Delta \omega$.

To derive the macroscopic energy functional $\mathcal{U}_{\text{tot}}$, we distinguish between the active pressure and the disorder potential for individual agents derived in Ref.~\cite{ivarsenPolarChiralActive2026}. First, the effective fluid self-interaction arises from the active stress tensor $\langle v_i v_j \rangle$ derived in Eq.~(\ref{eq:posit}). The isotropic component of this stress necessitates a macroscopic swim pressure that scales quartically with order,
\begin{equation} \label{eq:Pswim}
    P_\text{swim} \sim v_0^2 R^4.
\end{equation}
This equation of state ($P \propto \rho^2$, where $\rho=R^2$) corresponds to the barotropic potential energy density of a shallow water fluid.

We define the total energy functional $\mathcal{U}_{\text{tot}}[\rho]$ by combining this barotropic term with the topographic potential:
\begin{equation} \label{eq:shallow1}
    \mathcal{U}_\text{tot}[\rho] = \frac{g_{eff}}{2}(\rho - \rho_0)^2 + \gamma \rho \cdot \omega(\mathbf{r}),
\end{equation}
where $g_{\text{eff}}$ is the effective coupling constant representing the stiffness of the order parameter (analogous to gravity $g$ in shallow waters), and $\gamma$ is a coupling coefficient with dimensions $\text{length}^2 \text{time}^{-1}$ relating frequency disorder $\omega(\mathbf{r})$ to potential energy. Here, the frequency dispersion $\omega_i$, acting as a local conservative potential in the Josephson dynamics of the individual agents \cite{ivarsenPolarChiralActive2026}, maps to the external topographic potential $\omega(\mathbf{r})$ in the continuum limit (Figure~\ref{fig:topography}). We note that Eq.~(\ref{eq:shallow1}) corresponds directly to the potential energy of a fluid flowing over variable topography \cite{salmon_lectures_1998}. 

Two concise observations can now be made by dimensional analysis. \textit{(1)} The units of $\gamma$ naturally lead to the definition $\gamma=\sigma v_0$, anchoring the topographic potential to the physical interaction scale of the agents. \textit{(2)} By direct comparison between the divergence of the isotropic stress in Eq.~(\ref{eq:posit}) and the shallow water pressure gradient, we observe that $g_{\text{eff}} \equiv v_0^2/2$. This follows from observing that the active stress force $-\nabla \cdot (\rho \frac{v_0^2}{2})$ corresponds to the shallow water pressure force $-g_{\text{eff}}\nabla \rho$.

Next, the force per unit mass (acceleration) acting on a fluid element is the negative gradient of the variational derivative of the energy with respect to density \cite{salmon_lectures_1998}:
\begin{equation}
    \mathbf{F}_{\text{active}} = - \nabla \left( \frac{\delta \mathcal{U}_{\text{tot}}}{\delta \rho} \right).
\end{equation}
Substituting Eq.~(\ref{eq:shallow1}) yields,
\begin{equation} 
    \frac{\delta \mathcal{U}_{\text{tot}}}{\delta \rho} = \frac{v_0^2}{2}(\rho - \rho_0) + \sigma v_0 \omega(\mathbf{r}).
\end{equation}
Thus, the force field becomes,
\begin{equation}
    \mathbf{F}_{\text{active}} = - \underbrace{v_0^2 \nabla \rho/2}_{\text{Barotropic Pressure}} - \underbrace{\sigma v_0 \nabla \omega(\mathbf{r})}_{\text{Disorder Gradient}},
\end{equation}
where we note that $\nabla \omega$ appears as a conservative force field, akin to topographic steering in shallow waters \cite{salmon_lectures_1998}, and in adaptations to active nematics \cite{thijssen_submersed_2021}. Next, we will insert this force into the corrected Eulerian momentum equation for $\mathbf{u}$,
\begin{equation} \label{eq:Eumomentum}
    \frac{\partial \mathbf{u}}{\partial t} + (\mathbf{u} \cdot \nabla)\mathbf{u} = \mathbf{F}_{\text{active}} + \mathbf{F}_{\text{visc}},
\end{equation}
where we include a viscous force term $\mathbf{F}_{\text{visc}}$ for completeness, to be addressed later. Substituting $\rho = R^2$ and our derived force, we obtain,
\begin{equation} \label{eq:DuDt}
    \frac{D\mathbf{u}}{Dt} = - \frac{v_0^2}{2} \nabla (R^2) - \sigma v_0 \nabla \omega(\mathbf{r}) + \nu \nabla^2 \mathbf{u},
\end{equation}
where $\nu \nabla^2 \mathbf{u}$ is the viscous term, and where we identified the operator $D/Dt$ in Eq.~(\ref{eq:Eumomentum}), the material, or total, derivative \cite{kuramoto_chemical_2003}. This equation now mathematically represents a fluid flowing over a variable topography,
\begin{equation}
    Z(\mathbf{r}) = \frac{2\sigma}{v_0} \omega(\mathbf{r}),
\end{equation}
succinctly on display in Figure~\ref{fig:topography}. Since the inertial flow necessarily moves the distribution of $\omega_i$, the system is effectively advecting its own boundary conditions; the active particles produce the seafloor which provides topographic steering to the emergent inertial flow.

\subsection{Active acoustic speed}

At this point, we must address acoustics, as a shallow water vortex diffuses into the medium through sound waves. To demonstrate why our inviscid, dynamic dipole is unable to do so, we first revisit the continuity and momentum equations (Eqs.~\ref{eq:continuity} and \ref{eq:Eumomentum}) using $\rho = R^2$,
\begin{equation} \label{eq:cont2}
    \frac{\partial (R^2)}{\partial t} + \nabla \cdot (R^2 \mathbf{u}) = 0,
\end{equation}
and
\begin{equation}\label{eq:mom2}
    \frac{\partial \mathbf{u}}{\partial t} + (\mathbf{u} \cdot \nabla) \mathbf{u} = -\frac{v_0^2}{2} \nabla (R^2) - v_0\sigma \nabla \omega (\mathbf{r}).
\end{equation}
To find the speed of acoustic waves, we linearise the system around a base state of constant synchronisation $R_0(\mathbf{r})$ and zero velocity, neglecting the external topography $\omega(\mathbf{r})$ to isolate the acoustic modes, 
\begin{equation} \label{eq:Rdecomp}
    R(\mathbf{r}, t) = R_0(\mathbf{r}) +  \delta R(\mathbf{r}, t)
\end{equation}
and $\mathbf{u} = \mathbf{v}$ ($\mathbf{v}$ being a small perturbation). 
The linearised continuity and momentum equations then read,
\begin{equation}
     \frac{\partial \delta R(\mathbf{r}, t)}{\partial t} + \frac{R_0(\mathbf{r})}{2} \nabla \cdot \mathbf{v} = 0,
\end{equation}
and
\begin{equation}
     \frac{\partial \mathbf{v}}{\partial t} = -v_0^2 R_0(\mathbf{r}) \nabla \delta R(\mathbf{r}, t),
\end{equation}
respectively. Combining these by taking the time derivative of the continuity equation and substituting the momentum equation yields the wave equation:
\begin{equation}
    \frac{\partial^2 \delta R(\mathbf{r}, t)}{\partial t^2} - \frac{v_0^2 R_0(\mathbf{r})^2}{2} \nabla^2 \delta R(\mathbf{r}, t) = 0,
\end{equation}
from which we identify the active acoustic speed $c_s$,
\begin{equation} \label{eq:cs}
    c_s(R) = \frac{v_0 R_0(\mathbf{r})}{\sqrt{2}}.
\end{equation}

In other words, the sound-speed, or, in our case, propagation speed of oscillator state information, depends linearly on the local synchronisation $R$. When an acoustic wave travels from an ordered region ($R\approx1$) into a disordered region ($R\approx0$), a shock front necessarily forms, since the sound-speed rapidly decreases. The Mach number ($M$) of the flow, which is defined as the ratio of flow speed to sound speed, becomes,
\begin{equation} \label{eq:mach}
    M = \frac{|\mathbf{u}|}{c_s} \approx \frac{v_0 R}{v_0 R_0 / \sqrt{2}} = \sqrt{2} \left( 1 + \frac{\delta R}{R_0} \right),
\end{equation}
by merit of Eqs.~(\ref{eq:macro}) and (\ref{eq:cs}). The macroscopic flow is therefore inherently supersonic ($M > 1$). Near a defect, however, $R\to0$, and so $c_s\to0$; state information cannot escape a defect core, turning the latter into sonic black holes, aptly illustrated by the red `trapped' agents being effectively confined to the vortex cores in Figure~\ref{fig:0}. This implies the existence of a ``sonic horizon'' around defect cores, which become isolated behind Mach cones. Hence, defect cores cannot diffuse into their surroundings by radiating sound (such as geophysical vortices), justifying our application of the point-vortex Kirchhoff-Onsager Hamiltonian.

\begin{figure}
    \centering
    \includegraphics[width=0.5\textwidth]{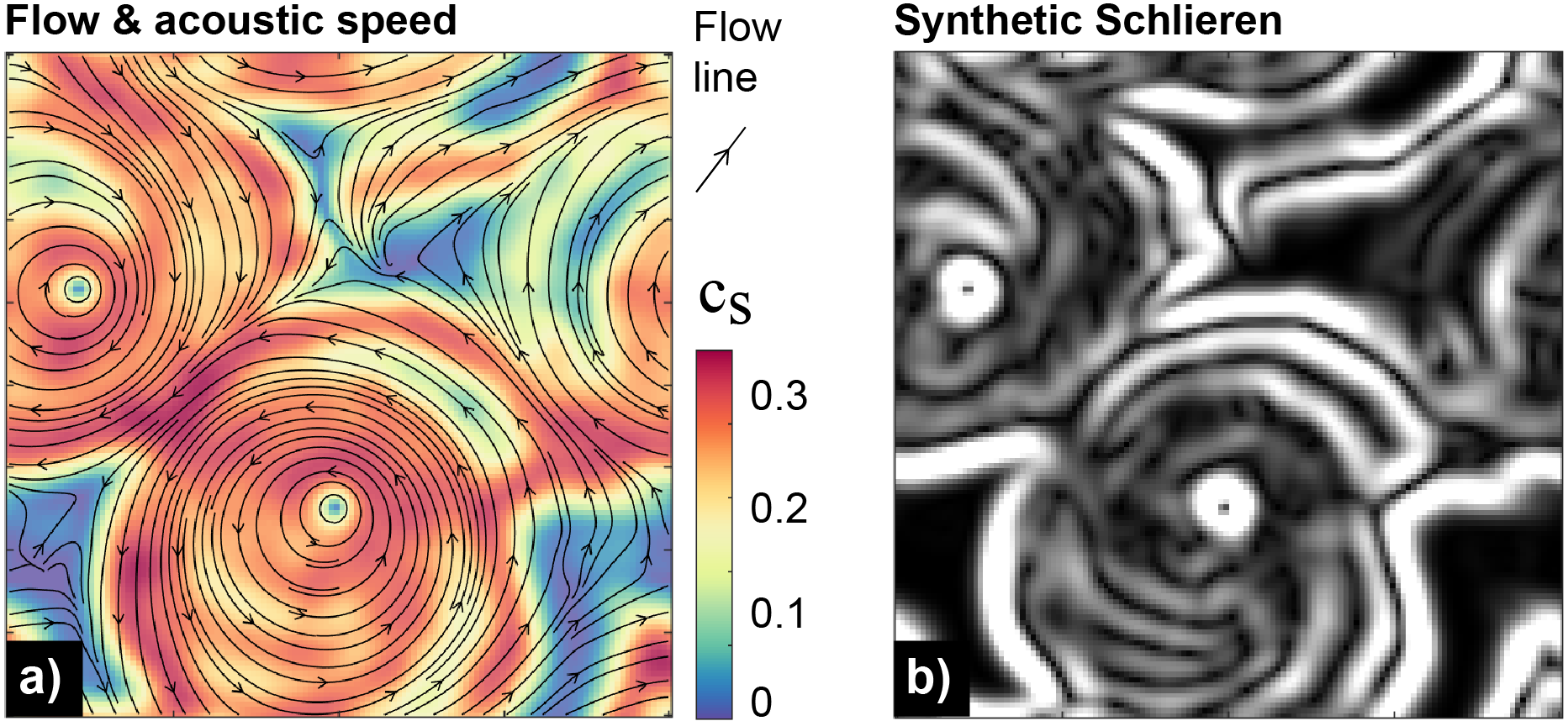}
    \caption{\textbf{Panel a)} shows acoustic speed (Eq.~\ref{eq:cs}) with a colormap, with $\mathbf{u}$ overlaid with flow lines, for an end-state simulation space. \textbf{Panel b)} shows a synthetic Schlieren plot ($|\nabla \rho|=|\nabla R^2|$), showing shock ridges.}
    \label{fig:shock}
\end{figure}

Figure~\ref{fig:shock} demonstrates the foregoing in empirical terms, showing the interplay between flow (panel a) and shock ridges (panel b) for the end-state of a Phase~III (``Onsager dipole'') simulation. We observe that \textit{(1)} the acoustic speed (colorscale in panel a) goes to zero near the defect cores, capturing the sonic black holes that are protected by Mach cones, and \textit{(2)} the synthetic schlieren plot \cite{settles_schlieren_2001} indicates the presence of shocks that radiate outward from the defect cores, where they thermalise inertial mass $R^2$.

We are now in a position to elucidate the dynamic behaviour observed in the hydrodynamic limit of our model. The vortex merging that we observe in Figures~\ref{fig:1} and \ref{fig:ensemble}, evidenced by the $t^{-0.75}$ scaling of the defect count, is explained by \textit{shock merging} in the hydrodynamic limit. That is, when two synchronisation shocks collide, $R$ decreases; the inertial energy contained in $R^2$ must be dissipated at the shock, going back into the active bath at the long-range strain interface between the defects. The active bath again synchronises, increasing $R$ and thereby creating inertial advection, sustaining the remarkably stable dynamics evident in Figure~\ref{fig:0} (see, also,  Appendix~A for additional numerical verification).  


To bolster the foregoing description, we return to Eq.~(\ref{eq:Rdecomp}) and note that $\delta R(\mathbf{r},t)$ is a fast, zero-mean fluctuation $\langle \delta R \rangle_t = 0$. We then linearise the density field $R^2$:
\begin{equation}
    R(\mathbf{r}, t)^2 = R_0(\mathbf{r})^2 + 2R_0(\mathbf{r}) \delta R(\mathbf{r}, t) + (\delta R(\mathbf{r}, t))^2.
\end{equation}
When we apply the time-average operator $\langle \cdot \rangle_t$ to this equation, the linear fluctuation term vanishes, and we are left with $\langle R^2 \rangle_t \approx R_0(\mathbf{r})^2$.

Next, we define the active velocity potential (or macroscopic phase) $\Theta(\mathbf{r},t)$ \cite{marchetti_hydrodynamics_2013},
\begin{equation}
    \Theta(\mathbf{r}, t) = \bar{\Theta}(\mathbf{r}, t) + \delta \Theta(\mathbf{r}, t),
\end{equation}
where a Helmholtz decomposition yields slow topological rotation (vortices) and the fast compressional vibration (acoustic waves). Our goal is to show that the slow-mode is incompressible, yielding the stable Euler dynamics that we observe in the main text. $\delta \Theta$ describes the irrotational phase oscillations associated with the sound waves $\delta R$, and likewise time-averages to zero: $\langle \delta \Theta \rangle_\tau = 0$. We substitute both expansions into the continuity equation,
\begin{equation}
    \partial_t R(\mathbf{r}, t)^2 + \nabla \cdot (R(\mathbf{r}, t)^2 \nabla \Theta(\mathbf{r}, t)) = 0,
\end{equation}
and time-average,
\begin{equation} \label{eq:monster}
    \nabla \cdot \langle (R_0(\mathbf{r})^2 + 2R_0(\mathbf{r}) \delta R(\mathbf{r}, t)) \nabla (\bar{\Theta}(\mathbf{r}, t) + \delta \Theta(\mathbf{r}, t)) \rangle_\tau \approx 0.
\end{equation}
After expanding Eq.~(\ref{eq:monster}), the second and third terms vanish by merit of the time-averaging, while the fourth term vanishes as $\delta R(\mathbf{r}, t)$ and $\nabla\delta\Theta(\mathbf{r}, t)$ are uncorrelated. We  are then left with the first dominant term, and recover the elliptic constraint for the vortex phase $\bar{\Theta}$ on the background $R_0$,
\begin{equation} \label{eq:rigidnew}
\nabla \cdot (R_0^2 \nabla \bar{\Theta}) = 0, 
\end{equation}
where we drop spatial and temporal dependencies for brevity. Eq.~(\ref{eq:rigidnew}) describes an incompressible potential flow in a medium with variable density $R_0^2$, and is analogous to the rigid lid approximation \cite{marchetti_hydrodynamics_2013}. We recognise a stable Euler fluid with inertial mass $R_0^2$, flowing around defect cores, and Eq.~(\ref{eq:rigidnew}) enforces mass conservation for the inertial background flow.

To address the physical nature of the background flow (which, at present, remains a static field approximation), we note that while the external topography $\omega(\mathbf{r})$ was neglected to isolate the fast acoustic modes ($\delta R$), it remains the governing factor for the slow background state $R_0(\mathbf{r})$. The static balance of the active enthalpy functional (Eq.~\ref{eq:shallow1}) dictates that $R_0(\mathbf{r})^2$ is spatially anti-correlated with $\omega(\mathbf{r})$, such that high-disorder regions create the density depressions (vacuums) that isolate defects, while low-disorder regions support the high-density ($R_0 \approx 1$) inertial fluid. A close examination of Figure~\ref{fig:topography}b, d) provides clear empirical evidence for this position.

\subsection{Hydrodynamic closure}

As the active acoustic speed (Eq.~\ref{eq:cs}) scales linearly with order ($c_s \propto R$),  
the sound speed inherently lags behind the inertial flow velocity by a constant, order unity, forcing the system into a global supersonic state with $M \approx \sqrt{2}$. As we discussed in the foregoing subsection, this condition mandates the formation of synchronisation shocks (Mach cones) at defect sites, which act as the primary mechanism of topological sorting (yielding Eulerian Kirchhoff-Onsager dynamics in the effective supersonic vortex fluid). This leads us to assert the rigid lid incompressibility,
\begin{equation}
    \nabla \cdot \mathbf{u} \approx 0.
\end{equation}
The ``phantom inertia'' $R^2$ acts as an Euler fluid with quantized circulation, as we demonstrate in the main text. At the same time, on fast, acoustic timescales, the dynamics are dominated by compressible shock-mergers. The shocks thermalise inertial energy into the active bath, which subsequently rebuilds order, feeding the inverse energy cascade. The rigid lid incompressibility emerges as the rigorous time-averaged limit of a never-ending stiff, supersonic topological gas.



%

\end{document}